\documentclass[fleqn,10pt]{wlscirep}
\usepackage[utf8]{inputenc}
\usepackage[T1]{fontenc}

\usepackage{hyperref}
\usepackage{url}
\usepackage{booktabs}
\usepackage{amsfonts}
\usepackage{nicefrac}
\usepackage{microtype}
\usepackage{xcolor}
\usepackage{graphicx}

\usepackage{amssymb}
\usepackage{siunitx}
\usepackage{enumitem}
\usepackage{float}

\usepackage{eucal}
\usepackage{multirow}

\usepackage{marginnote}

\usepackage{comment}
\usepackage{authblk}

\begin{document}
\definecolor{DarkGreen}{RGB}{0, 100, 0}

\makeatletter 
\@mparswitchfalse 
\makeatother

\hypersetup{
    colorlinks=true,
    linkcolor=blue,
    filecolor=magenta,      
    urlcolor=cyan,
    citecolor=blue,
}

\title{QSAR-Guided Generative Framework for the Discovery of Synthetically Viable Odorants}

\author[1]{Tim C. Pearce\thanks{Corresponding author: t.c.pearce@le.ac.uk}}
\author[2]{Ahmed Ibrahim}

\affil[1]{Biomedical Engineering Research Group, School of Engineering, University of Leicester, University Road, Leicester LE1 7RH, United Kingdom}
\affil[2]{School of Clinical Medicine, University of Cambridge, Addenbrooke's Hospital, Hills Road, Cambridge CB2 0SP, United Kingdom}

\date{}

\begin{abstract}
The discovery of novel odorant molecules is key for the fragrance and flavor industries, yet efficiently navigating the vast chemical space to identify structures with desirable olfactory properties remains a significant challenge. Generative artificial intelligence offers a promising approach for \textit{de novo} molecular design but typically requires large sets of molecules to learn from. To address this problem, we present a framework combining a variational autoencoder (VAE) with a quantitative structure-activity relationship (QSAR) model to generate novel odorants from limited training sets of odor molecules. The self-supervised learning capabilities of the VAE allow it to learn SMILES grammar from ChEMBL database, while its training objective is augmented with a loss term derived from an external QSAR model to structure the latent representation according to odor probability. While the VAE demonstrated high internal consistency in learning the QSAR supervision signal, validation against an external, unseen ground truth dataset (Unique Good Scents) confirms the model generates syntactically valid structures (100\% validity achieved via rejection sampling) and 93.9\% unique structures. The latent space is effectively structured by odor likelihood, evidenced by a Fréchet ChemNet Distance (FCD) of $\approx$ 6.96 between generated molecules and known odorants, compared to $\approx$ 21.6 for the ChEMBL baseline. An ablation of the property head shows that its contribution lies in the density of retrievable structures near an odorant seed --- a 4.3-fold increase in distinct novel compounds per seed --- rather than in the odor-likeness of any individual compound. Bemis-Murcko analysis shows 2,499 of 6,765 ring-containing candidates, spanning
1,429 distinct core frameworks, absent from all training sets, while a further 294 molecules,
spanning 57 frameworks, independently recover odorant cores found only in an external catalog
withheld from training. Generated candidates display physicochemical properties consistent with ground-truth odorants (mean MW $\sim$159 Da, LogP $\sim$1.67) and comparable predicted ADMET profiles. Furthermore, quantum mechanical calculations (GFN2-xTB) verify thermodynamic stability with energy distributions matching known volatiles, and automated retrosynthesis identifies candidate synthesis routes, yielding valid synthesis routes for 100\% of candidates, averaging 2.89 steps from commercially available precursors. This integrated approach provides a novel and systematic methodology for applying generative AI to explore chemical space specifically for the discovery of new candidate odorant molecules.
\end{abstract}

\flushbottom
\maketitle

\begin{figure}[H]
  \centering
  \includegraphics[width=0.7\textwidth]{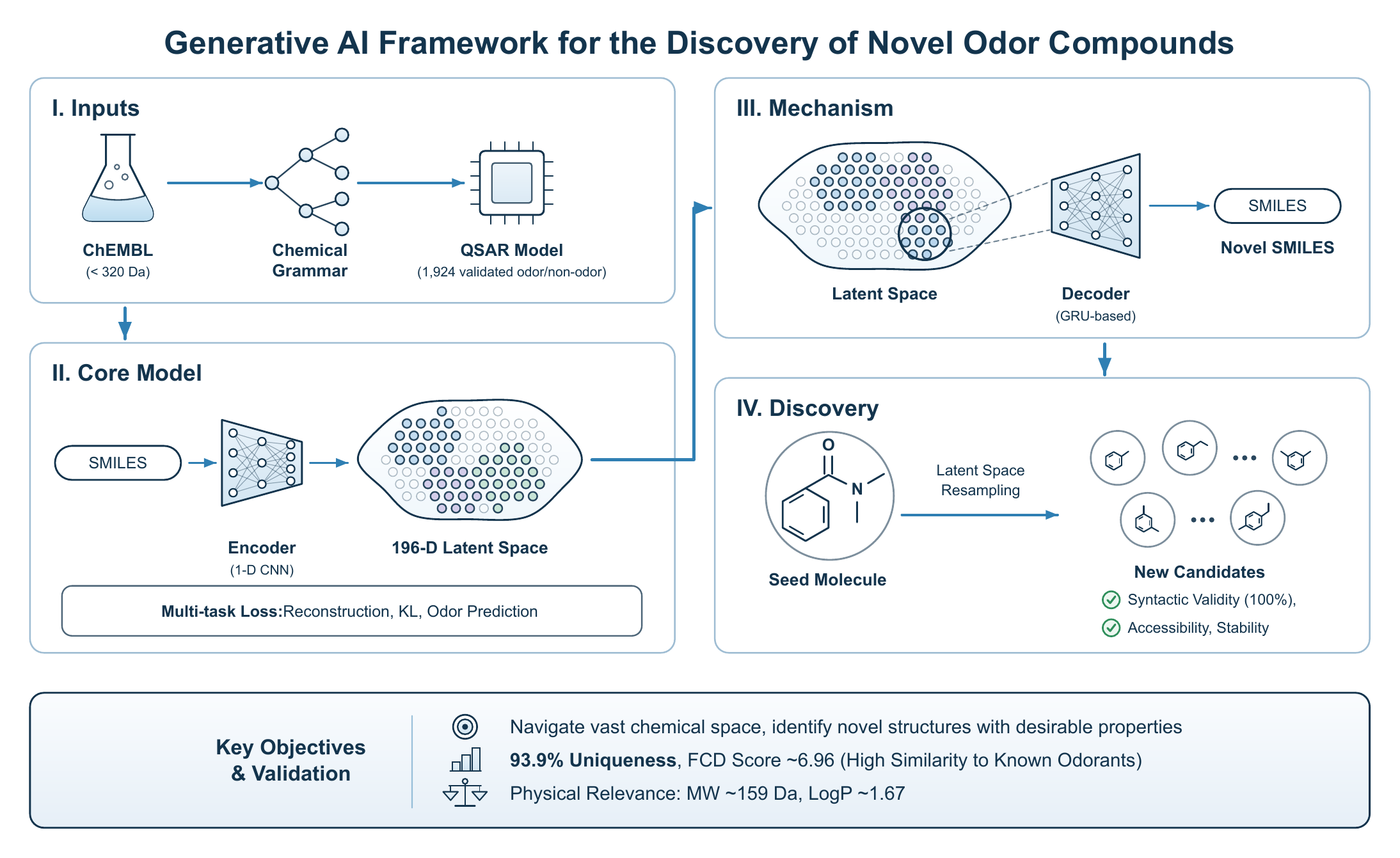}
\end{figure}

\textbf{Keywords:} odorant molecules, olfaction, computational chemistry, molecule design, generative AI

\section*{Introduction}
Olfactory perception (whether orthonasal or retronasal) fundamentally shapes consumer experience, making the discovery of new scents and flavors a key competitive advantage and an area of significant research investment [1]. This is a dynamic interdisciplinary endeavor, driven by the
innovation needs of the fragrance and food industries, commercial demands for unique and cost-effective ingredients, and increasingly stringent regulatory constraints regarding safety and
environmental impact, and a growing imperative for sustainable and ethically sourced materials
[2]. These factors necessitate a constant search not only for replacements for restricted
ingredients but also for molecules offering enhanced performance, such as improved stability in
challenging product matrices, novel olfactory characteristics that create new trends, and added
consumer benefits like improved diffusion, longevity, or malodor counteraction [3].
Consequently, there is substantial interest in understanding structure-odor relationships (SORs) and developing new approaches for odor molecule discovery and design [4].

Understanding the link between molecular structure and perceived odor presents a formidable
scientific challenge. Molecules must possess specific physicochemical properties: sufficient
volatility, moderate hydrophobicity, and typically a molecular weight below \textasciitilde\SI{300}{\dalton} to reach and
interact with olfactory receptors (ORs) in the nasal epithelium [5]. The human olfactory system
employs \textasciitilde400 functional ORs in a complex combinatorial coding strategy, where single receptors
recognize multiple odorants and single odorants activate multiple receptors [6]. This system
produces a vast range of distinct perceived odors – while the exact number is debated, the
potential perceptual odor space is immense, generated from a chemical space potentially
spanning billions of discrete molecules [7]. The highly dimensional molecular and perceptual
space is sensitive to subtle structural nuances; therefore, discovering new molecules with
diverse structures is essential for industrial applications while also advancing our fundamental
understanding of the olfactory pathway and odor perception.

Traditionally, novel odorant discovery relies heavily on synthetic organic chemistry, often guided
by intuition and empirical observation based on known natural or synthetic odorants [1].
Chemists systematically modify structures or explore derivatives of readily available chemical feedstocks, iteratively synthesizing analogs and evaluating their scent organoleptically. While
successful, this `molecule-first' approach can be laborious and may struggle to efficiently
navigate the vast potential chemical space to find truly novel structures meeting specific
performance, property or safety criteria.

Recent advances in computational methods, particularly generative artificial intelligence (AI),
offer powerful new tools to address these challenges. Over the past five years, techniques like
variational autoencoders (VAEs), generative adversarial networks (GANs), transformers, and
diffusion models have profoundly impacted molecular discovery, enabling rapid exploration of
chemical space and \textit{de novo} design [8–10]. These methods facilitate `inverse design', shifting
focus from modifying known structures to generating entirely new molecules optimized \textit{a priori} to
satisfy key property profiles [11]. By using desired characteristics as inputs, AI models propose
candidates meeting complex multi-objective criteria. However, while this property-driven approach has shown
success in yielding experimentally validated molecules for pharmaceutical targets [12], its application
to the specific domain of olfaction remains comparatively limited. The discovery of novel odorants imposes
uniquely stringent requirements—necessitating precise physicochemical properties for volatility alongside
rigorous dermatological and environmental safety compliance—that generic molecular generators often fail
to capture. Consequently, the field holds significant promise for accelerating odorant innovation only
when these generative capabilities are tightly integrated with strict property checks and predictions of
synthetic feasibility to bridge the gap between theoretical design and practical synthesis [13].

Generative AI approaches generally rely on large labeled datasets, which are often unavailable for specific properties like odor [14]. Directly training generative models on the limited sets of known odorants is insufficient to capture the complex syntax of SMILES or establish a continuous, interpolatable latent space. This paper introduces a framework to overcome this limitation. We combine a predictive quantitative structure-activity relationship (QSAR) model,
trained on a curated set of known odorants, with a molecular variational autoencoder (VAE). This
combined generative framework synergizes the representational power of a VAE with the
predictive capability of a QSAR model tailored for odorant properties. The approach leverages the
VAE's ability to learn the underlying `grammar' of molecular structures from large, general
chemical databases (ChEMBL) via self-supervision, while using the QSAR model's predictions to
specifically guide the VAE's learning process and structure its latent space towards olfactory
relevance.

The subsequent sections detail this integrated methodology: the Methods outlines the VAE and QSAR model architectures and the combined training strategy; the Results presents the validation of the QSAR model and the VAE's generative performance, with a primary focus on validating the physicochemical, structural, and electronic properties of the generated compounds against the generic chemical baseline (ChEMBL) rather than benchmarking generative architectures; and the Perspective and conclusions discusses the implications of this guided generative approach for accelerating the discovery of next-generation fragrance and flavor ingredients.

\section*{Methods}
\subsection*{Molecular variational autoencoder architecture}
The variational autoencoder architecture consists of four main components: an encoder, a latent
space, an odor prediction head and a decoder. We base the VAE architecture on that reported in
[15], adding a QSAR-driven odor prediction head (Figure 1).

\begin{figure}[h!]
  \centering
  \includegraphics[width=1.0\textwidth]{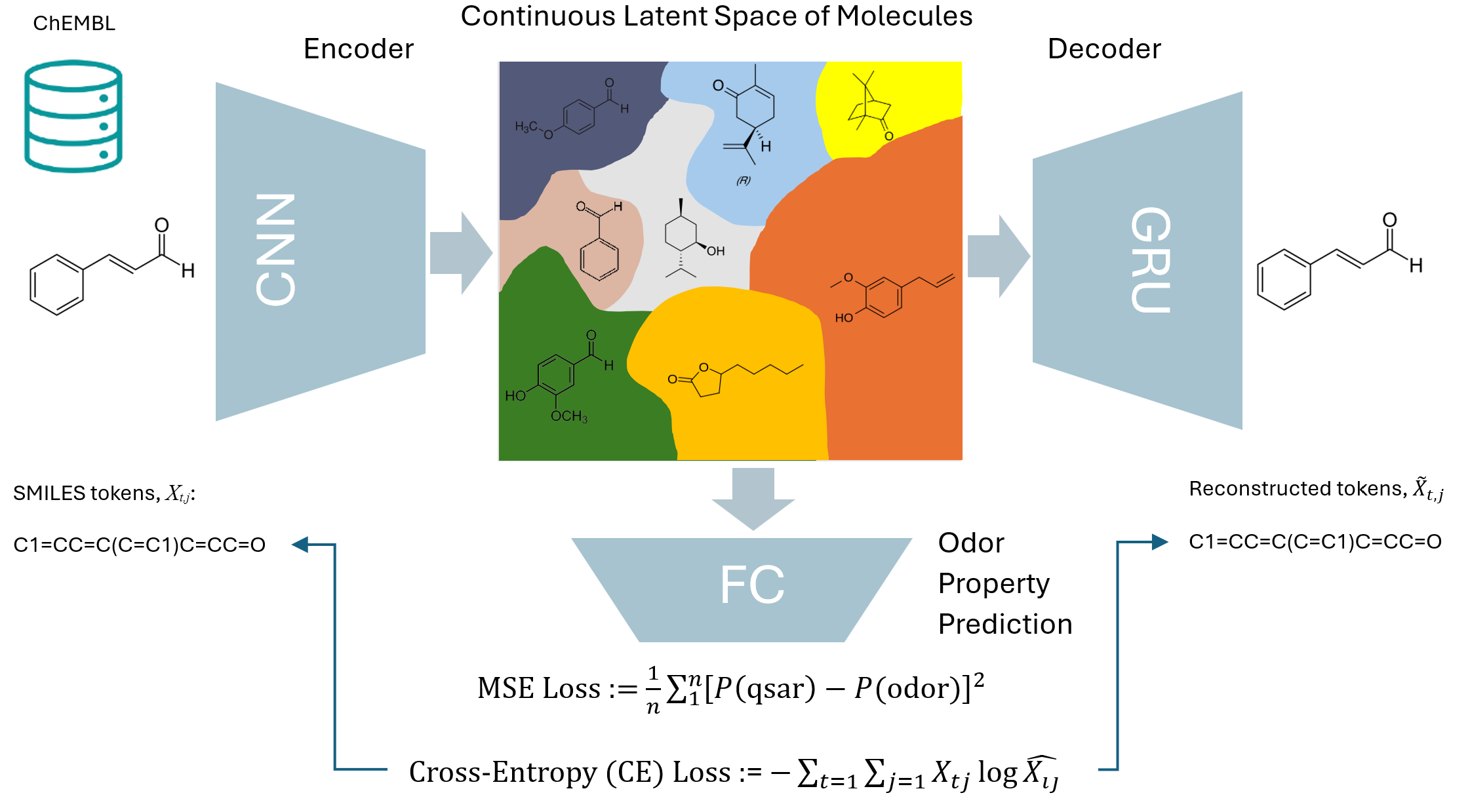}
  \caption{Variational Autoencoder framework concept. Abbreviations: CNN, convolutional neural
network; GRU, gated recurrent unit; FC, fully connected (dense); KL, Kullback-Leibler divergence;
Recon Loss, reconstruction loss (categorical cross-entropy); Prop Loss, property prediction loss
(mean squared error, as described in text). The diagram illustrates the flow: input SMILES from
ChEMBL are processed by an Encoder (CNN based) mapping to a continuous latent space. Latent
vectors $z$ are input to both a Decoder (GRU based) for reconstructing SMILES sequences and an
Odor Property Prediction head (FC based) predicting odor probability $P(\text{odor})$ jointly with two auxiliary regression targets (LogP and boiling point), of which only $P(\text{odor})$ enters the property loss. The combined
training objective minimizes the Recon Loss, the Prop Loss, and the KL divergence regularization
term.}
  \label{fig:fig1}
\end{figure}

\paragraph{SMILES sequence embedding:}
A single input SMILES (Simplified Molecular Input Line Entry System) sequence is padded to a
fixed length $T$, resulting in a character sequence $s' = (c_1, c_2, ..., c_T)$, where each
character $c_t$ belongs to the predefined vocabulary $\mathcal{C}$ of size $V$. This fixed length $T$ is determined by the maximum sequence length observed across the entire training dataset to ensure no input molecule's representation is truncated. To represent this sequence
numerically, one-hot encoding is performed. First, we define a mapping $\text{index}: \mathcal{C} \rightarrow \{0,1, ..., V-1\}$ that assigns a unique integer index to each character in the vocabulary. Each
character $c_t$ in $s'$ is then transformed into a one-hot vector $v_t \in \{0,1\}^V$, such that the $j$-th
component of this vector (where $j$ ranges from 0 to $V-1$) is determined by comparing $j$ with the
unique index assigned to the character $c_t$, using the Kronecker delta
\begin{equation}
(v_t)_j = \delta_{j,\text{index}(c_t)}, \label{eq:1}
\end{equation}
where $\delta_{a,b}$ equals 1 if $a = b$ and 0 otherwise.

These one-hot vectors are arranged into a matrix $X$ of shape $T \times V$, where each row corresponds
to a character token in the sequence. The scalar element $X_{t,j}$ at row $t$ and column $j$ of this
matrix $X$ is therefore given by
\begin{equation}
X_{t,j} = \delta_{j,\text{index}(c_t)} \quad \text{for } t \in \{1, ..., T\}, j \in \{0, ..., V-1\} . \label{eq:2}
\end{equation}
This means that for each row $t$ of the matrix $X$, only the single element in the
column $j$ corresponding to the index of character $c_t$ will be 1, and all other elements in that row
will be 0. This matrix $X \in \{0,1\}^{T \times V}$ serves as the input tensor to the VAE encoder.

\paragraph{VAE encoder:}
The encoder network maps an input SMILES token sequence to the parameters of a latent
Gaussian distribution. Each SMILES (represented by input tensor $X$) is directly processed by the
first 1-D convolution layer. In this context, the sequence length $T$ serves as the steps (or length)
dimension along which the 1-D convolution kernel slides. The vocabulary size $V$ acts as the
channels dimension; at each time step $t$, the channels map onto the elements of the $V$-dimensional one-hot vector within $X$. Therefore, the 1-D convolution layer applies its filters
across a window of time steps, considering all $V$ channels within that window to compute its
output features. The notation $h^{(0)}$ used in the description below simply refers to this initial $T \times V$ input tensor $X$ before any convolutional transformations are applied.

The encoder is structured as a series of convolutional layers, designed to extract meaningful
features from the input molecular representation. Each subsequent convolutional layer increases in both depth (number of filters per layer) and width (size of each kernel) according to growth factors (optimized to  1.16 for depth and 1.18 for
width), enabling the network to capture increasingly complex molecular features at different scales for learning hierarchical representations of molecular structures.  More formally, the input $h^{(0)}$ (as layer 0) first passes through a sequence of three one-dimensional convolutional layers ($j = 1 ... 3$).

The kernel sizes are chosen to capture common, local chemical motifs within that span such as
aromatic rings and functional groups of the SMILES sequence (layer 1: 9 filters, kernel size 9, layer
2: 9 filters, kernel size 9, layer 3: 10 filters, kernel size 11). Subsequent layers progressively
increase the number of filters by a geometric factor, allowing the network to learn a richer
hierarchy of features by combining simpler patterns, while the stacking of layers increases the
effective receptive field to capture longer-range dependencies. Each layer applies the
convolution operation (although implemented as cross-correlation in Keras because flipping the
kernel is an unnecessary computational step when the kernel's weights are learned during
training) on the previous layer's output as follows
\begin{equation}
h_{\text{conv}}^{(j)} = \tanh (W_{\text{conv}}^{(j)} * h^{(j-1)} + b_{\text{conv}}^{(j)}), \label{eq:3}
\end{equation}
using a hyperbolic tangent activation, followed by batch normalization
\begin{equation}
h_{\text{norm}}^{(j)} = \text{BatchNorm} (h_{\text{conv}}^{(j)}) = \gamma^{(j)} \frac{h_{\text{conv}}^{(j)}-\mu_{\mathcal{B}}}{\sqrt{\sigma_{\mathcal{B}}^2 + \epsilon}} + \beta^{(j)}, \label{eq:4}
\end{equation}
to stabilize training. Here $\mu_{\mathcal{B}}$ is the mini-batch mean, $\sigma_{\mathcal{B}}^2$ is the mini-batch variance, $\epsilon$ is a small
constant added to the variance for numerical stability ($10^{-5}$) to prevent division by zero in case the
mini-batch variance is zero, $\gamma^{(j)}$ is a learnable scaling parameter allowing the network to adjust
the variance of the normalized activations, and $\beta^{(j)}$ is a learnable shifting parameter that allows
the network to adjust the mean of the normalized activations. Batch normalization was applied
after each convolutional layer to improve generalization by normalizing the inputs to each layer.

The output tensor from the final convolutional layer is flattened into a vector $h^{(4)}$. This vector is
processed by a single dense layer employing tanh activation and batch normalization, with
dropout applied, allowing for mixing of the extracted convolutional features. Let the output of this
layer be denoted $h_{\text{enc}}$, which represents a compressed encoding of the input SMILES sequence.

The parameters of the latent Gaussian distribution $q_{\phi}(z|X) \sim \mathcal{N}(\mu_{\phi}, \sigma_{\phi}^2 I)$ are then derived from
this intermediate encoding $h_{\text{enc}}$ in two distinct steps, handled by different parts of the model
structure:
\begin{enumerate}
    \item \textbf{Latent mean ($\mu$) vector calculation:} The main encoder function computes the latent mean vector $\mu$ by applying a dense layer with a linear activation function directly to the intermediate encoding $h_{\text{enc}}$
    \begin{equation}
    \mu_{\phi} = W_{\mu}h_{\text{enc}} + b_{\mu}. \label{eq:5}
    \end{equation}
    The encoder model function returns both this calculated mean $\mu$ and the intermediate encoding $h_{\text{enc}}$.
    \item \textbf{Latent log-variance ($\log \sigma^2$) vector calculation:} A separate function also takes the intermediate encoding $h_{\text{enc}}$ as input. It then applies a separate dense layer with a linear activation function, to $h_{\text{enc}}$ to compute the latent log-variance vector $\log \sigma_{\phi}^2$
    \begin{equation}
    \log \sigma_{\phi}^2 = W_{\sigma^2}h_{\text{enc}} + b_{\sigma^2}. \label{eq:6}
    \end{equation}
\end{enumerate}
To enable loss gradient backpropagation through the sampling process a reparameterization trick
expresses the latent vector $z$ as a deterministic function of the learned mean $\mu_{\phi}$, log-
variance $\log \sigma_{\phi}^2$ and independent standard Gaussian noise, $\varepsilon$ [16]. Specifically, $z$ is computed
as $z = \mu_{\phi} + \exp(0.5 \log \sigma_{\phi}^2) \odot \varepsilon$, where $\varepsilon \sim \mathcal{N}(0, I)$. This isolates the randomness in $\varepsilon$, allowing
gradients to flow from the VAE loss back to the encoder parameters $\mu$ and $\log \sigma^2$ for end-to-end
training.

The dimensionality of the latent space, $D$, defines the capacity of the variational bottleneck. This
separation allows the main encoder model structure to focus on producing the mean and the
feature representation, while the dedicated variational layers function handles the variance
prediction and the sampling process.

\paragraph{VAE latent space:}
The encoder network, parameterized by $\phi$, maps an input SMILES token sequence $X$ onto a
probabilistic representation in the latent space. Specifically, it outputs the parameters of an
approximate posterior distribution $q_{\phi}(z|x)$ as a multivariate Gaussian with a diagonal covariance
matrix
\begin{equation}
q_{\phi}(z|X) = \mathcal{N}(z|\mu_{\phi}(X),\text{diag}[\sigma_{\phi}^2(X)]). \label{eq:7}
\end{equation}
Here, $z \in \mathbb{R}^D$ is the latent vector corresponding to the input $X$ governed by $q_{\phi}(z|X)$, which defines
the location and spread in the $D$-dimensional latent space. Therefore, instead of a deterministic
point, multiple latent vectors $z$ may be obtained for each SMILES $X$ by sampling from this
distribution.

\paragraph{VAE decoder:}
The decoder network reconstructs the SMILES sequence from a single sampled latent vector $z \in \mathbb{R}^D$. The sampled vector first passes through a single dense transformation layer. This layer
employs a hyperbolic tangent (tanh) activation function followed by batch normalization, to
prepare the latent representation for sequential processing
\begin{equation}
h_{\text{dec}}^{(1)} = \text{BatchNorm} (\tanh (W_{\text{dec}}^{(1)} z + b_{\text{dec}}^{(1)})), \label{eq:8}
\end{equation}
with dropout applied. The resulting vector $h_{\text{dec}}^{(1)}$ is then replicated $T$ times (the maximum
sequence length) using a repeat vector operation, creating an input sequence $H_{\text{repeat}} = [h_{\text{dec}}^{(1)}, h_{\text{dec}}^{(1)}, ..., h_{\text{dec}}^{(1)}]$ for the recurrent part of the decoder.

This sequence is processed by a stack of three standard Gated Recurrent Units (GRUs) ($k = 1,2,3$), chosen for their effectiveness in capturing temporal dependencies while mitigating
vanishing gradient issues. Each standard GRU layer $k$ computes its hidden state, $h_t^{(k)} \in \mathbb{R}^{488}$,
based on the previous hidden state, $h_{t-1}^{(k)}$, and the output, $h_t^{(k-1)}$from the layer below ($h_t^{(0)}$ being
the $t$-th vector from $H_{\text{repeat}}$).

The GRU update equations are standard:
\begin{subequations}
\begin{align}
r_t^{(k)} &= \sigma_g (W_r^{(k)}h_t^{(k-1)} + U_r^{(k)}h_{t-1}^{(k)} + b_r^{(k)}), \label{eq:9a}\\
u_t^{(k)} &= \sigma_g (W_u^{(k)}h_t^{(k-1)} + U_u^{(k)}h_{t-1}^{(k)} + b_u^{(k)}), \label{eq:9b}\\
\tilde{h}_t^{(k)} &= \tanh (W_h^{(k)}h_t^{(k-1)} + U_h^{(k)}(r_t^{(k)} \odot h_{t-1}^{(k)}) + b_h^{(k)}), \label{eq:9c}\\
h_t^{(k)} &= (1 - u_t^{(k)}) \odot h_{t-1}^{(k)} + u_t^{(k)} \odot \tilde{h}_t^{(k)}, \label{eq:9d}
\end{align}
\label{eq:9all}%
\end{subequations}
where $\sigma_g$ is the sigmoid function, tanh is the internal activation and $\odot$ is
element-wise multiplication. Stacking GRUs allows the decoder to model increasingly complex
temporal patterns necessary for generating syntactically valid SMILES sequences.

The final layer ($k = 4$) is a custom Terminal GRU (TGRU) layer designed to leverage teacher forcing
during training. In this mode, the TGRU's state update incorporates the true previous token from
the input sequence, $x_{t-1}^{\text{true}}$ (the one-hot vector for the ground truth character at step $t-1$), in
addition to the input from the layer below ($h_t^{(3)}$) and its own previous hidden state ($h_{t-1}^{(4)}$).
Specifically, the candidate hidden state $\tilde{h}_t^{(4)}$ calculation is modified to include a term dependent
on the true previous input, gated by the reset gate $r_t^{(4)}$
\begin{equation}
\tilde{h}_t^{(4)} = \tanh (W_h^{(4)}h_t^{(3)} + r_t^{(4)} \odot (U_h^{(4)}h_{t-1}^{(4)} + U_x^{(4)} x_{t-1}^{\text{true}}) + b_h^{(4)}), \label{eq:10}
\end{equation}
where $U_x^{(4)}$ represents the weights applied specifically to the teacher-forced input $x_{t-1}^{\text{true}}$. This
teacher forcing mechanism provides the network with the correct preceding character during
training, which can stabilize learning and accelerate convergence for sequence generation tasks. This resulting hidden state $h_t^{(4)}$ from the TGRU encapsulates the information needed to predict
the probability distribution for the next token.
The TGRU is configured with $V$ units, so that the projection to the
vocabulary dimension producing the raw, unnormalized scores (logits) $o_t \in \mathbb{R}^V$,
$o_t = W_{\text{out}}h_t^{(4)} + b_{\text{out}}$, is performed by the output kernel of the
recurrent cell itself rather than by a separate dense layer; $W_{\text{out}}$ and
$b_{\text{out}}$ are its learnable weight matrix and bias vector.
The softmax activation of that same cell, moderated by a
temperature $\tau$, converts these logits into the reconstructed probability distribution vector $\hat{X}_t$ over the vocabulary $V$
for the token at time step $t$
\begin{equation}
\hat{X}_{t,i} = \frac{\exp(o_{t,i}/\tau)}{\sum_{k'=1}^V \exp(o_{t,k'}/\tau)}, \label{eq:11}
\end{equation}
where $\hat{X}_{t,i}$ is the $i$-th element of the vector $\hat{X}_t$, representing the probability assigned to the $i$-th
vocabulary character at time step $t$. The temperature parameter $\tau$ specifically modifies the
character sampling process used during sequence generation (e.g., at test time), adjusting the
trade-off between exploiting high-probability tokens (lower $\tau$) and exploring more diverse options
(higher $\tau$). The default value $\tau = 1.0$ results in sampling based on the unmodified output
probabilities during generation.

\paragraph{Odor prediction head:}
For any decoded SMILES token sequence the combined model features a dedicated property
prediction head that takes the latent vector $z$ as input. This regression head begins with a dense
layer compressing the latent representation for the decoded molecule into a 67-dimensional
space, initiating a processing pathway focused on property feature prediction. Two additional
dense layers are applied to add depth to the property pathway. The dimensionality decreases only marginally through these layers
(67, 66 and 65 units respectively; a ratio of $0.99$ between consecutive layers). Batch normalization
is applied after each intermediate layer, which helps stabilize training dynamics and dropout is
applied to improve generalization. The network culminates in a specialized output layer: one
employing a linear activation function to predict continuous values suitable for regression tasks
(optimized using Mean Squared Error – see Section below).The final layer of the odor prediction head is a linear layer with three
regression outputs, corresponding to the jointly predicted properties $P(\text{odor})$,
LogP and boiling point. Only the first of these, $P(\text{odor})$, enters the property loss
of Equation~\ref{eq:16} and is reported throughout this work; the two auxiliary targets act
as additional supervision on the latent representation. 

\subsection*{Model training}
The combined VAE-QSAR model is trained end-to-end by optimizing the parameters of the
encoder ($\phi$), odor prediction head and decoder ($\theta$) to maximize the Evidence Lower Bound (ELBO)
on the marginal log-likelihood of the data $\log p (X)$. This is equivalent to minimizing the negative
ELBO, which serves as the primary loss function. The loss function $\mathcal{L_{\text{VAE}}}$ for a single input
sequence $X$ is composed of three terms: a reconstruction loss, a Kullback-Leibler (KL) divergence
regularization term and an odor property prediction loss.
\begin{equation}
\mathcal{L}_{\text{VAE}}(X, \theta, \phi) := \underbrace{\beta_{KL}(e) \cdot D_{KL}(q_{\phi}(z|X) \| p(z))}_{\text{Regularization (KL Divergence)}} + \underbrace{\mathcal{L}_{\text{recon}}}_{\text{Reconstruction Loss}} + \underbrace{\gamma \mathcal{L}_{P(\text{odor})}}_{\text{Property Prediction Loss}} \label{eq:12}
\end{equation}
Here, $q_{\phi}(z|X)$ is the approximate posterior distribution $\mathcal{N}(z|\mu_{\phi}(X),\text{diag}[\sigma_{\phi}^2(X)])$ parameterized
by the encoder's outputs ($\mu_{\phi}, \log \sigma_{\phi}^2$), $p_{\theta}(X|z)$ is the likelihood of the data given the latent
variable parameterized by the decoder, and $p(z)$ is the prior distribution over the latent variables,
chosen as a standard multivariate Gaussian $\mathcal{N}(z|0, I)$.

The reconstruction loss, $L_{\text{recon}}$, measures how accurately the decoder reconstructs the input
using the latent representation of $X$ alone. Given the categorical nature of SMILES tokens and the
decoder's softmax output $\hat{X}_{t,i} = p_{\theta}(X_t = i|z)$, this term is implemented as the categorical cross-entropy, summed over the sequence length $T$ and vocabulary size $V$
\begin{equation}
\mathcal{L_{\text{recon}}} := - \sum_{t=1}^T \sum_{i=1}^V X_{t,i} \log(\hat{X}_{t,i}). \label{eq:13}
\end{equation}
The base weight for this reconstruction term is set to $w_{\text{xent}} = 1.0$.

The KL divergence term acts as a regularizer, encouraging the approximate posterior $q_{\phi}(z|x)$ to
stay close to the prior $p(z)$. For the chosen Gaussian distribution, it has an analytical form
calculated over the latent dimension
\begin{equation}
D_{KL}(q_{\phi}(z|X)\|p(z)) = \frac{1}{2}\sum_{j=1}^D(\sigma_{\phi,j}^2 + \mu_{\phi,j}^2 - 1 - \log\sigma_{\phi,j}^2). \label{eq:14}
\end{equation}
The weight $\beta_{KL}(e)$ applied to the KL divergence term is annealed during training scheduled over
epoch $e$. This annealing begins with the KL weight near zero and gradually increases it, allowing
the model to initially focus on achieving good reconstruction before strongly enforcing the latent
space structure. The annealing follows a cyclical schedule centered at epoch $e_{\text{start}} = 22$ with a
slope $s = 1.0$, scaled by a target KL weight $\beta_{\text{target}} = 1$
\begin{equation}
\beta_{KL}(e) = \beta_{\text{target}} \frac{1}{1+\exp(-s(e-e_{\text{start}}))}. \label{eq:15}
\end{equation}
This annealing strategy helps prevent the KL term from collapsing the latent space prematurely
("posterior collapse").

\begin{table}[h]
\centering
\caption{Hyperparameter Optimization Search Space. The optimal values selected for the final VAE-QSAR model are highlighted in \textbf{bold}.}
\label{tab:hyperparameters}
\begin{tabular}{lccccc}
\toprule
\textbf{Hyperparameter} & \multicolumn{5}{c}{\textbf{Values Evaluated}} \\
\midrule
Latent Space Dimensions ($D$) & 64 & 128 & \textbf{196} & 256 & 512 \\
GRU Hidden Units & 128 & \textbf{488} & 512 & 768 & 1024 \\
Convolutional Growth Factor & 1.00 & 1.10 & 1.12 & \textbf{1.16} & 1.25 \\
Learning Rate ($\eta$) & $1.0 \times 10^{-3}$ & $5.0 \times 10^{-4}$ & \textbf{3.12 $\times$ 10$^{-4}$} & $1.0 \times 10^{-4}$ & $1.0 \times 10^{-5}$ \\
Adam Optimizer Decay ($\beta_1$) & 0.900 & 0.925 & \textbf{0.937} & 0.950 & 0.990 \\
Batch Size ($B$) & 32 & 64 & \textbf{100} & 128 & 256 \\
KL Annealing Center ($e_{\text{start}}$) & 10 & 15 & \textbf{22} & 30 & 50 \\
Odor Head Compression Ratio & 0.50 & 0.65 & 0.75 & 0.80 & \textbf{0.99} \\
CNN Kernel Size (Layer 1) & 3 & 5 & 7 & \textbf{9} & 11 \\
Softmax Temperature ($\tau$) & 0.5 & 0.8 & \textbf{1.0} & 1.2 & 1.5 \\
\bottomrule
\end{tabular}
\end{table}

The overall loss, averaged over a mini-batch of size $B$, is minimized using the Adam optimizer.
Adam is chosen for its adaptive learning rate capabilities, with a learning rate $\eta = 3.12 \times 10^{-4}$ and a first moment estimate decay factor $\beta_1 = 0.937$. To ensure the robustness of the generative framework, we performed a systematic grid search across the key architectural and optimization hyperparameters using the candidate values listed in Table \ref{tab:hyperparameters}. The final configuration (highlighted in bold) was selected based on performance metrics evaluated on the validation set, specifically prioritizing parameter combinations that minimized the overall ELBO loss while maintaining stable convergence and high reconstruction accuracy. Training was performed for a specified number of epochs using 10\% (validation
split = 0.1) of the data for validation to monitor generalization and prevent overfitting. The entire
training procedure aims to find model parameters ($\theta, \phi$) that yield both accurate reconstructions
and a well-structured latent space adhering to the Gaussian prior.
The odor prediction head is trained using the probability of the corresponding molecule being
odorous according to the QSAR model, $P(\text{qsar})$, following a mean-square loss (MSE) criterion over
a mini-batch (see the section ``QSAR odor prediction model'' below on the QSAR logistic regression model used to generate
these targets). For a single SMILES sequence
\begin{equation}
\mathcal{L_{P(\text{odor})}}:=[P(\text{qsar})-P(\text{odor})]^2. \label{eq:16}
\end{equation}
During training, the reconstruction and KL terms are weighted equally
($w_{\text{xent}} = \beta_{\text{target}} = 1$), with the property prediction term weighted by
$\gamma = 0.5$. Balancing these multiple objectives means that the VAE's latent space
possesses several key properties for encoding molecular structures. The space is shaped
by regularization and smoothness constraints imposed by the KL divergence term in the VAE
objective. This term encourages the learned distributions $q_{\phi}(z|X)$ to stay close to a prior $p(z)$,
forcing encodings into a structured region and promoting continuity where similar molecules
map to nearby, overlapping distributions, enabling meaningful interpolation for molecular
generation. Lastly, the dimensionality (optimization settled on $D = 196$) of the latent space
controls the balance between compactness and expressiveness. It dictates the number of
dimensions available to capture molecular variations, with smaller $D$ promoting efficiency but
potentially limiting detail, while larger $D$ allows richer representation but requires sufficient
regularization to remain meaningful.

The batch size $B = 100$
was found to balance computational efficiency and the stochasticity needed for effective training.
The number of training epochs is determined based on convergence and acceptable validation
set overall loss. The encoder's convolutional stack along with subsequent layers that further
expand geometrically using these factors, aims to capture diverse local chemical motifs and
progressively build hierarchical features with an increasing receptive field. Standard stride
1 and `valid' padding are used within the CNN layers. 
The recurrent component employs four recurrent layers --- three standard GRUs
with a recurrent dimension of 488, followed by a TGRU whose units are of the vocabulary
dimension $V$.
Training used
the Adam optimizer, selected for its adaptive capabilities, optimized to a learning rate $\eta = 3.12 \times 10^{-4}$ and $\beta_1 = 0.937$. Overall, these parameters were found to capture hierarchical sequence
features accurately, learning a structured latent space via controlled regularization, and
achieving stable convergence.

\subsection*{Scaffold classification}
Generated molecules were classified by Bemis-Murcko scaffold through a hierarchical cascade
against three reference sets, applied in the order: the QSAR training set, the ChEMBL corpus
used for VAE training, and the external Unique Good Scents catalog. Each molecule is assigned
to the first category it satisfies, the five categories being those of
Table~\ref{tab:scaffold_analysis}: structural match to a reference molecule (Exact
Memorization); scaffold present in the QSAR training set (Odorant Derivatization); scaffold
present in ChEMBL (Repurposing); scaffold present in Unique Good Scents but in neither training
set (Validated Scaffold Hop); scaffold absent from all three (Uncharted Scaffold Hop).
The QSAR training set vocabulary comprises the Bemis-Murcko scaffolds of all $1{,}924$ training
molecules ($1{,}865$ distinct structures, $264$ distinct scaffolds), odor- and
odorless-labelled alike, since the comparison is structural rather than label-conditioned. Six
of the $3{,}291$ category-2 molecules ($0.18\%$) match a scaffold occurring only among the
odorless training compounds.
RDKit returns an empty scaffold for any molecule containing no ring system. Acyclic compounds
therefore have no framework that can be matched against a reference vocabulary, and the
framework-based categories (2 to 5) are undefined for them. Such compounds are excluded from the
scaffold classification, which is reported over the ring-containing subset; by construction,
categories 2 to 4 contain no acyclic compounds. Because the acyclic fraction of the library is
substantial, it is assessed separately by nearest-neighbor Tanimoto similarity on ECFP4
fingerprints (radius 2, 2048 bits). Supplementary Table~S2
reports the Uncharted Scaffold Hop rate under each of the three possible conventions for
handling acyclic compounds.

\paragraph{Ablation of the odor prediction head:}
To establish that the QSAR-driven property loss (the term $\gamma\mathcal{L}_{P(\text{odor})}$ in
Equation~\ref{eq:12}) is itself responsible for organizing the latent space by olfactory relevance,
we additionally trained an ablated model in which that term is removed
($\gamma\mathcal{L}_{P(\text{odor})}\!=\!0$), leaving a conventional VAE trained on reconstruction
and KL regularization alone. The ablated model is otherwise identical to the full model: the same
convolutional encoder, the same $D=196$ latent space, the same GRU/TGRU decoder, the same 
ChEMBL
 training corpus, and the same 300 training epochs. In the ablated configuration the reconstruction
and KL weights were both rescaled by a common factor ($w_{\text{xent}}\!=\!\beta_{\text{target}}\!=\!0.25$),
preserving their 1:1 ratio; since the Adam update is invariant to a uniform scaling of the loss,
this leaves the reconstruction--KL balance --- the quantity that governs latent geometry ---
unchanged. We note further that a down-weighted reconstruction term would, if anything, be expected
to \emph{degrade} decoder fidelity, whereas the ablated model in fact reconstructs markedly
\emph{better} than the full model (Section ``Ablation control''); the direction of that difference
therefore excludes the loss rescaling as an explanation of the results reported there. The
motivation for the ablation, and the generative comparison performed on it, are given in the
Results.

\begin{figure}[!htp]
  \centering
  \includegraphics[width=0.8\textwidth]{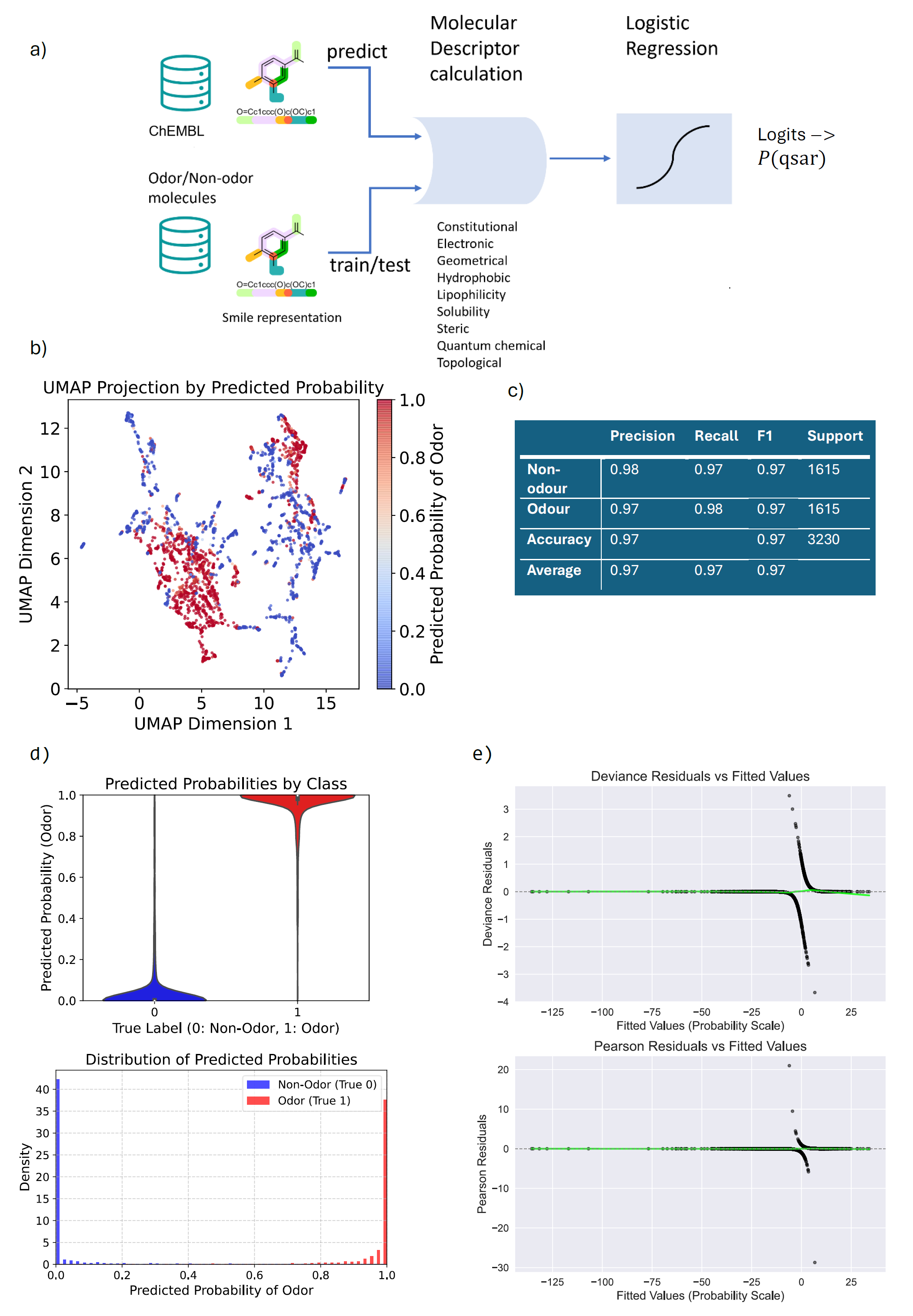}
  \caption[QSAR Odor molecule prediction results.]{QSAR Odor molecule prediction results. a) Schematic showing the QSAR
model pipeline where logistic regression was trained on 1,924 experimentally validated and/or
literature reported molecules and subsequently used to label ChEMBL molecules with a target
odor probability, b) UMAP visualization of training set in space of molecular descriptors showing
the odor probability, c) classifier performance for logistic model, and d) Distribution of predicted probabilities across the
training set, and e) residual plots for the logistic regression model.}
  \label{fig:fig2}
\end{figure}

\subsection*{QSAR odor prediction model}

Molecular descriptors $m^{(i)}$ for each molecule $i$ were calculated using Mordred descriptor
calculator [17] covering a wide set of properties (Figure 2a). Both 2-D and 3-D descriptors were
used, resulting in a total of 1,828 molecular features. Descriptors that generated non-numeric
results for at least one training set molecule were discarded. Highly correlated ($\rho > 0.99$)
descriptors were removed. Descriptors with low variance across the data-set ($\sigma^2 < 0.05$ after
autoscaling) were also removed, resulting in 370 remaining descriptors used for the analysis.

To generate the targets for the odor prediction head of the VAE we trained a logistic regression
model on a previously validated set of 1,615 odor and 309 non-odor molecules [14]. The SMOTE algorithm was used to correct this class imbalance by oversampling the minority (non-odor) class, giving $n = 1,615$ for both classes [18]. The endpoint of this model is odorousness, that is, whether a molecule is perceptible as
an odorant at all; it does not predict odor quality, intensity or detection threshold. The
underlying labels [14] derive from transport features and do not distinguish activation of
olfactory receptors from trigeminal chemesthesis, so a compound predicted to be odorous may be
perceived as an irritant rather than as an odor.

For a vector of molecular descriptors $m^{(i)}$ corresponding to a given molecule $i$, we used a logistic
regression model that describes the probability of this molecule being odorous in terms of a
single binary-dependent variable $y^{(i)}$
\begin{equation}
P(\text{qsar}) = P(y^{(i)} = 1|m^{(i)}) = f_{\theta}(m^{(i)}) = \frac{1}{1+e^{-\theta^T m^{(i)}}}, \label{eq:17}
\end{equation}
where $\theta$ denotes the parameter vector to be learnt (often termed logistic regression beta
coefficients) across all molecular descriptors. To optimize this, we minimize the logistic loss
(negative log-likelihood) cost function over a training set of $n$ molecules, which is equivalent to
maximizing the likelihood of observing the given data under the logistic regression model
\begin{equation}
J(\theta) = - \frac{1}{n} \sum_{i=1}^n [y^{(i)} \log(f_{\theta}(m^{(i)})) + (1 - y^{(i)}) \log (1-f_{\theta}(m^{(i)}))], \label{eq:18}
\end{equation}
where $y^{(i)}$ is the class label (0 non-odor and 1 odor) and $m^{(i)}$ is the feature vector for the $i^{th}$
training example (molecule in training set).

Loss minimization is performed using gradient descent with the following update rule for each
parameter within $\theta$
\begin{equation}
\theta_j := \theta_j-\alpha \frac{1}{n}\sum_{i=1}^n(f_{\theta}(m^{(i)})-y^{(i)})m_j^{(i)}, \label{eq:19}
\end{equation}
where $\alpha$ is the learning rate. Regularization strength was selected by five-fold cross-validation within the
balanced training set; because oversampling preceded the fold split, these folds are not fully
independent, and the external Good Scents evaluation reported in the Results provides the
unbiased performance estimate.

\begin{figure}[!htp]
  \centering
  \includegraphics[width=1.0\textwidth]{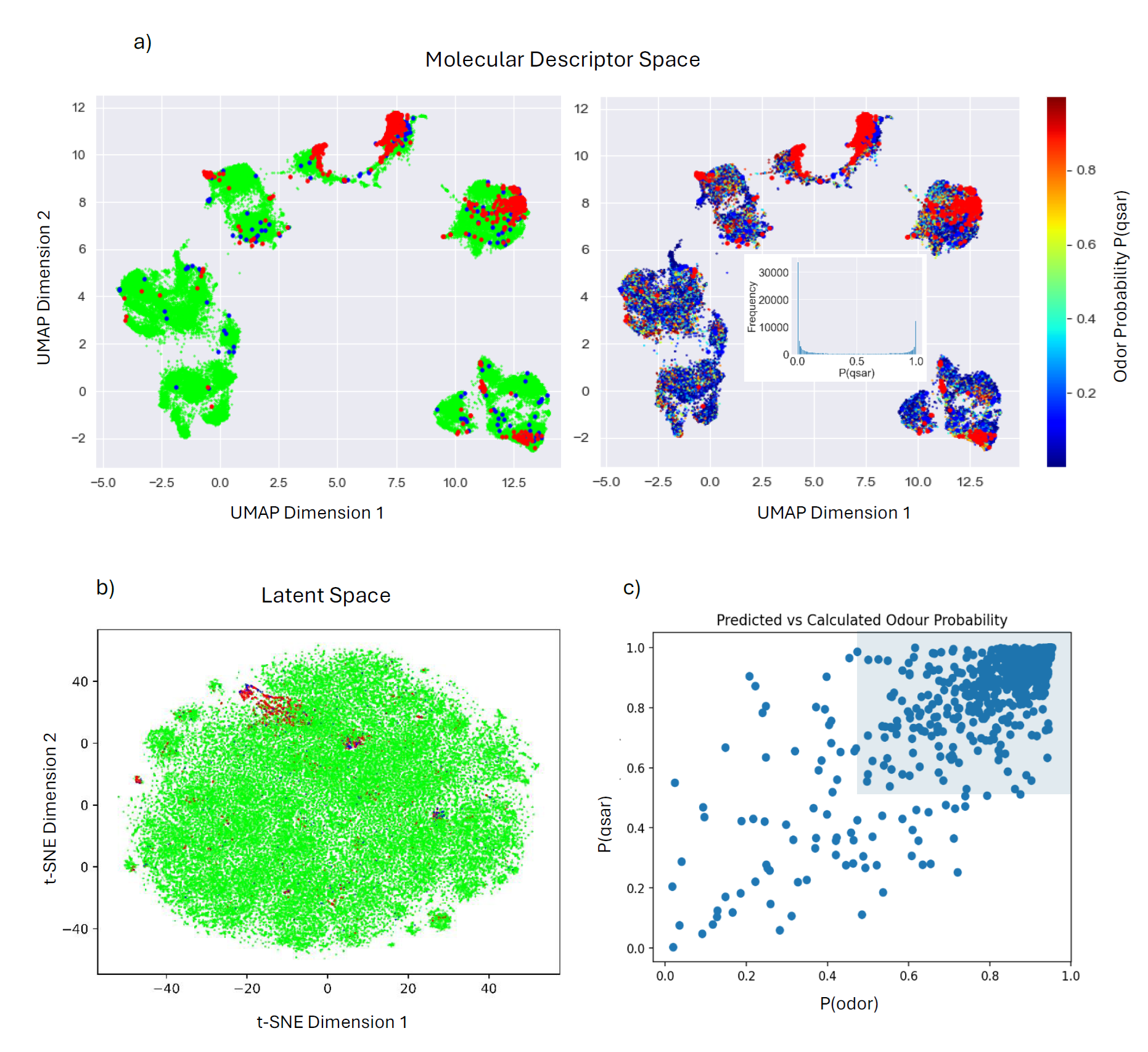}
  \caption[UMAP and t-SNE dimensionality reduction of the molecular descriptor and latent spaces.]{a) UMAP dimensionality reduction and visualization of the molecular descriptor space
showing \textasciitilde5 x $10^5$ VAE training set ChEMBL molecules (green) alongside the original set of odor
molecules (red) and non-odor molecules (blue) and the same UMAP representation showing the
predicted probabilities across all ChEMBL molecules used for VAE training as predicted by the
QSAR model (right). Inset histogram for the frequency of odor probabilities in the ChEMBL set
used to train the VAE, b) t-SNE dimensionality reduction of 1 x $10^4$ randomly sampled ChEMBL
molecules encoded into the latent space (green) together with validated odor (red) and non-odor
molecules (blue), c) Validated odor molecule probabilities as predicted by the QSAR model,
$P(\text{qsar})$, compared against VAE odor property prediction, $P(\text{odor})$. Shaded region shows where
both models agree on odor molecule prediction ($P(\text{odor}) > 0.5$ and $P(\text{qsar}) > 0.5$).}
  \label{fig:fig3}
\end{figure}

\subsection*{Applicability domain}
The applicability domain of the QSAR model was defined by nearest-neighbor Tanimoto similarity
on ECFP4 fingerprints (radius 2, 2048 bits) between each query molecule and the QSAR training
set. A reference threshold was derived from the training set itself, as the 5th percentile of
its own leave-one-out nearest-neighbor distribution ($0.2465$; leave-one-out mean $0.560$,
median $0.542$). Because domain membership is strongly threshold-dependent, it is reported at
three thresholds rather than as a single figure. Where a single criterion is required for
candidate selection we use $0.35$, a round value lying between the training-derived threshold
of $0.2465$ and $0.40$, and report membership at all three.

\paragraph{Reactive chemotype screening:}
Generated structures were screened for reactive
functional groups using SMARTS substructure matching in RDKit. The patterns applied identify acyl
halides, sulfonyl halides, phosphoryl halides, isocyanates and carboxylic anhydrides. These five patterns are applied directly; no
published structural alert catalogue was used. Matching
compounds are excluded from the proposed candidate set. Isothiocyanates were deliberately not
screened, being represented among cataloged odorants. The SMARTS definitions are given in the
Supplementary Information and in the deposited code.

\section*{Results}
\subsection*{Odor prediction QSAR model validation}
Fitted to the SMOTE-balanced QSAR training set, the logistic regression model
returned precision, recall and F1 of 0.97 with an overall accuracy of 97\% (Figure 2c); these are
resubstitution figures on the data to which the model was fitted. The five-fold cross-validation
performed internally during regularization selection gives a mean accuracy of $0.929 \pm 0.028$
across folds. As an external test, the fitted model was applied to the `Unique Good Scents'
catalog \protect\footnote{\url{https://www.thegoodscentscompany.com/allodor.html}}, which was unseen during both QSAR and VAE training and consists entirely of catalogued
odorants: of 4,343 compounds, 4,337 could be described by the retained descriptors, and 89.1\% of
these were classified as odorous ($P(\text{qsar})>0.5$; mean $P(\text{qsar})=0.875$, median
$0.997$). The model therefore achieves an external sensitivity of 89.1\% on odorants it had never
seen.
Predicted odor probabilities separated
adequately showing acceptable error for use in subsequent training of the VAE (mean probability
for non-odor class = 0.043, odor class = 0.957, Figure 2d). UMAP was used on the 370 descriptors
for dimensionality reduction and visualization of predicted probabilities in the molecular
descriptor space (Figure 2b).

\begin{table}[h]
  \centering
  \begin{tabular}{ll}
    \toprule
    Dep. Variable: & odor.class \\
    No. Observations & 3230 \\
    Model & Logit \\
    Df Residuals & 2858 \\
    Method & MLE \\
    Df Model & 371 \\
    Pseudo R$^2$ & 0.894 \\
    Log-Likelihood & -235.80 \\
    LL-Null & -2238.9 \\
    LLR p-value & 0.000 \\
    \bottomrule
  \end{tabular}
\caption{Logistic Regression Model Diagnostics}
\label{tab:logit_diagnostics}
\end{table}

After fitting the logistic function, diagnostics and classification performance were checked
(Table~\ref{tab:logit_diagnostics}). Logistic regression is based on the following assumptions that were tested on the fitted model
[19]:
\begin{enumerate}
    \item Binary response variable: Logistic regression assumes that the response variable has
    only two possible outcomes – verified odor and non-odor molecules were applied in the
    training set so only dichotomous training data were considered.
    \item Independence of errors: Logistic regression assumes that errors (residuals) are
    independent. Figure 2e shows the residual errors for the training set in the original and
    Pearson standardized coordinates have no repeat error values (monotonic).
    \item Observations are independent: Logistic regression requires that the observations be
    independent of each other, meaning they should not come from repeated measurements
    or matched data. As each molecule is separately sampled across a random set of training
    molecules, observations can be considered independent.
    \item No multicollinearity: Logistic regression requires that the independent variables are not
    too highly correlated with each other. This is ensured by removing highly correlated
    molecular descriptors before logistic regression is applied.
\end{enumerate}
The rank order of the absolute model coefficient values showing the contribution of each Mordred
molecular descriptor to predicting odor probability is shown in Figure S1 (descriptor names can
be resolved from [20]).

\begin{figure}[tp]
  \centering
  \includegraphics[width=0.8\textwidth]{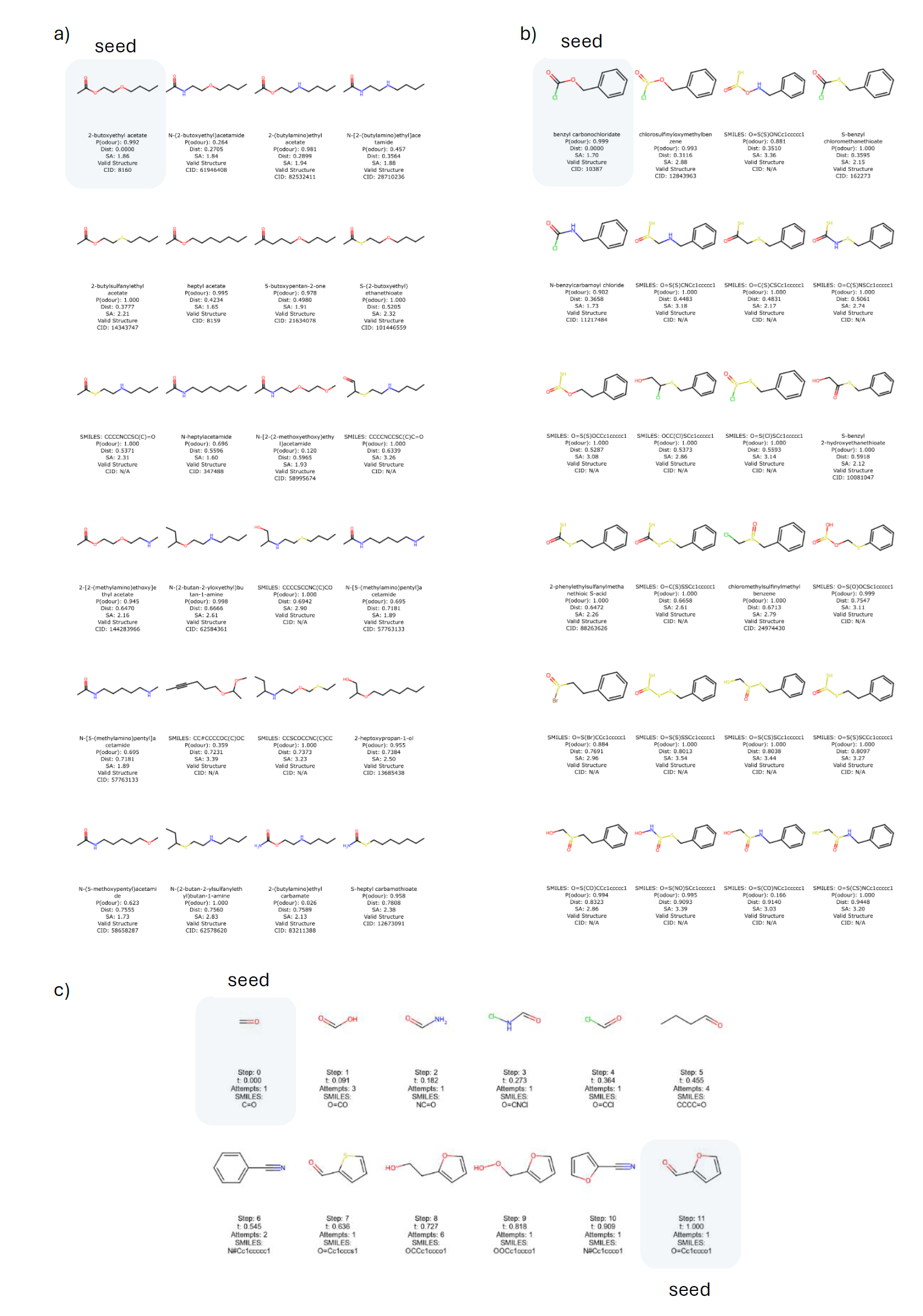}
  \caption{Generated molecules from two example seed odor molecules. a \& b) Two example seed
molecules (seed is distance 0.0, top left) with sampling of the latent space over increasing
distance from the seed in latent space (units of $\sigma_{\phi}(z|X)$). Where available Pubchem catalogued
compounds show their corresponding IUPAC name and CID reference, whereas potentially novel
compounds show the corresponding SMILES and CID: N/A. SA: synthesis accessibility score. c)
traversing the latent space between two odor seed molecules $t = 0$ is start seed and $t = 1$ is the
end seed in 11 discrete steps. Attempts are the number of probabilistic decode cycles required
to obtain a valid structure at that point of the latent space.}
  \label{fig:fig4}
\end{figure}

\subsection*{VAE training with ChEMBL molecules}
To provide a sufficiently large set of molecules for VAE training, the fitted QSAR model was applied
to a random sample of \textasciitilde5 x $10^5$ compounds (< \SI{320}{\dalton}) from ChEMBL{} [21] to predict their
probability of odor, $P(\text{qsar})$. The molecular descriptor space was then visualized using UMAP
against the original QSAR validated odor/non-odor set (Figure 3a). ChEMBL molecules were
chosen due to bioactive properties but also a low occurrence of odor molecules - to specifically
test the hypothesis that the final generated molecules still retain odor characteristics in contrast
to the training set (see histogram inset Figure 3a, right). Importantly, the validated odor/non-odor
molecule set was not used to train the VAE directly, but rather via ChEMBL to provide sufficient
examples to learn the SMILES grammar itself and QSAR model generated target probabilities.

After training the VAE on the ChEMBL molecule set we examined their representation within the
196-dimensional latent space (Figure 3b) using t-SNE and plotted them alongside the original
odor/non-odor molecule set. With some exceptions, QSAR training set odor compounds (red)
were found to cluster within the overall latent space distinct from the non-odor and ChEMBL
molecules. This suggests that the multi-task loss function is effective in organizing the latent
space representation by odor properties even though the VAE was not trained on the QSAR training
odor molecule set directly.

To check the fidelity of the knowledge distillation process from QSAR to VAE models, we compared the $P(\text{odor})$ values from the VAE odor prediction head  against the corresponding QSAR model predicted values, $P(\text{qsar})$ (Figure 3c) using the original training set. The shaded region shows classification agreement between the two models, yielding a precision of 0.98, a recall of 0.97, and an F1-score of 0.97. This high agreement confirms that the VAE successfully captured the QSAR model's decision boundary and encoded the supervisory signal into its latent space. However, as the QSAR model serves as the training objective, this agreement represents internal consistency rather than external biological validation that comes later. The rigorous test of the model's ability to generalize to true olfactory biology is presented in subsequent sections using the external `Unique Good Scents' dataset, which was unseen by both the QSAR and VAE components.

\subsection*{Generative molecule resampling from the latent space}
Because the VAE maps SMILES onto a probability distribution with mean and log variance within
the latent space, $q_{\phi}(z|X)$, resampling is possible to generate closely related decoded molecules.

To generate new molecules near a known seed molecule in the latent space, the VAE's encoder
first predicts the mean latent representation ($\mu_{\text{seed}}$) from the corresponding seed SMILES
tensor $X_{\text{seed}}$, which is then standardized using the global mean ($\mu_{\text{global}}$) and standard deviation
($\sigma_{\text{global}}$; both of which are computed across the training set's latent space distribution during
training), such that $z_{\text{seed}} = (\mu_{\text{seed}} - \mu_{\text{global}})/\sigma_{\text{global}}$. To explore the latent space around this
standardized point, a resampling process adds controlled noise. Briefly, a noise vector ($\varepsilon_{\text{noise}}$) is
sampled from a standard multivariate normal distribution $\mathcal{N} (0, I)$, normalized to unit magnitude,
$\hat{\varepsilon}_{\text{noise}}$, and then scaled by a specified scalar distance parameter, $d$, which dictates the Euclidean
distance of the perturbation in the latent space. This scaled noise vector ($\delta z = d\hat{\varepsilon}_{\text{noise}}$) is added
to the standardized latent seed point: $z_{\text{perturbed}} = z_{\text{seed}} + \delta z$. Before this perturbed vector can
be processed by the decoder, it must be unstandardized back to the original scale of the latent
space using the inverse transformation: $z_{\text{dec}} = z_{\text{perturbed}}\odot\sigma_{\text{global}} + \mu_{\text{global}}$.
Finally, the decoder network takes this resulting
unstandardized vector as input and translates it into a new candidate SMILES string $X_{\text{gen}}$. To ensure the production of physically interpretable structures, a rejection sampling strategy was employed. If the decoded SMILES sequence failed syntax validation checks (via RDKit [22]), the decoding process was repeated probabilistically for that latent point until a valid structure was obtained or a maximum retry limit was reached (see Figure 4c for attempt counts)

To test whether the odor properties of generated molecules are locally related in latent space we
then applied as input the QSAR training set as seed molecules to the combined model and
resampled the latent space at varying distances from the seeds (Figure 4a,b). To assess their odor likelihood  we then ran these generated molecules through the fitted QSAR model.
The majority of generated molecules are found to have $P(\text{qsar})$ and $P(\text{odor}) > 0.9$ (Figures 4, 6a (inset)). SMILES sequence validity was checked using
standard RDKit syntax check and sanitization (chemical plausibility check).

\paragraph{Generation workflow:}
The seed set comprises the $n=1{,}615$ odor-classified compounds of the QSAR training set [14]; non-odorant compounds were not used as seeds. Each seed was canonicalized, encoded to its latent mean $\mu_{\text{seed}}$, and decoded $N=100$ times. To explore the neighborhood, a fresh random Gaussian perturbation scaled by a batch noise norm of $10.0$ (corresponding to an average Euclidean distance of $0.5\sigma$ in standardized latent space) was drawn for each decode attempt. Decoded strings failing RDKit syntax and sanitization checks were rejected (rejection sampling), and surviving structures were canonicalized and de-duplicated within each seed. Across the full seed set, this yielded $37{,}163$ structures ($33{,}507$ distinct after canonicalization), averaging approximately $23$ distinct compounds per seed, which constitutes the library analyzed throughout the remainder of this work.

\subsection*{Molecular VAE encoding capacity}
The calculated average pairwise Euclidean distance between seed molecules in the latent space
of $2.67 \pm 0.79$ (units of $\sigma_{\phi}(z|X)$) provides a quantitative measure of the encoding capacity of the
combined model when compared with distances of generated molecules (Figure 4a,b). The ablation control reported below was performed on a separately retrained full model, for which the corresponding mean inter-seed distance is $2.59$. The inter-seed distance compared against the distances of resampled points in the latent space shows at
least one order of magnitude encoding capacity greater than seed compounds originally used for
training, which is essential for discovering novel chemical entities. Depending on search
distances explored from seeds we found each run generates between 20 – 50k unique molecules
indicating an encoding capacity of at least 10 - 25 times multiplier on the seed odor set.

As a follow-up study, generated molecules and their corresponding $P(\text{qsar})$ values could be
reintroduced for self-supervised training to further improve the encoding capacity of the
combined model and more fully explore the latent space embeddings.

\subsection*{Molecular diversity and odor relevance}
The generated molecules are found to contain diverse structural features and functional groups known to be
important in odor and flavors and many compounds belong to multiple categories (Supplementary Data).
\begin{itemize}
    \item Aromatic compounds: many molecules contain benzene rings, indicated by notations like \texttt{`c1ccccc1'}. Examples include benzaldehyde (\texttt{`O=Cc1ccccc1'}) and various substituted benzenes.
    \item Aldehydes: molecules with the \texttt{`C=O'} group, such as \texttt{`O=Cc1ccccc1'}.
    \item Carboxylic acids: compounds ending with \texttt{`C(=O)O'}, such as \texttt{`CC(=O)O'}.
    \item Ketones: containing a carbonyl group between carbon atoms, e.g., \texttt{`CC(=O)CC(C)C'}.
    \item Alcohols: compounds containing the \texttt{`-OH'} group, such as \texttt{`CC(=O)CC(C)O'}.
    \item Amines: molecules with nitrogen atoms, like \texttt{`CCNC1CCCCCC1'}.
    \item Heterocyclic compounds: cyclic structures containing atoms other than carbon, e.g., \texttt{`O=C1CCCCO1'}.
    \item Alkenes: molecules with carbon-carbon double bonds, like \texttt{`C=C(C)C'}.
    \item Halogenated compounds: molecules containing halogens (F, Cl, Br), such as \texttt{`CCC(C)F'}.
    \item Sulfur-containing compounds: molecules with sulfur atoms, e.g., \texttt{`CCCCCSCCCC'}.
    \item Esters: compounds with the \texttt{`COC=O'} group, like \texttt{`CC(=O)OC(C)C'}.
    \item Nitriles: molecules containing the \texttt{`C\#N'} group, such as \texttt{`N\#Cc1ccccc1'}.
    \item Ethers: compounds with the \texttt{`C-O-C'} linkage, e.g., \texttt{`COCCCCOCC'}.
    \item Thiols and sulfides: molecules containing \texttt{`S-H'} or \texttt{`C-S-C'} groups, like \texttt{`CSCCS'}.
    \item Terpenes and terpenoids: complex molecules often found in essential oils, like \texttt{`CC(C)=CCCC(C)(O)C1CC=C(C)CC1'}.
\end{itemize}

\begin{figure}[tp]
  \centering
  \includegraphics[width=1\textwidth]{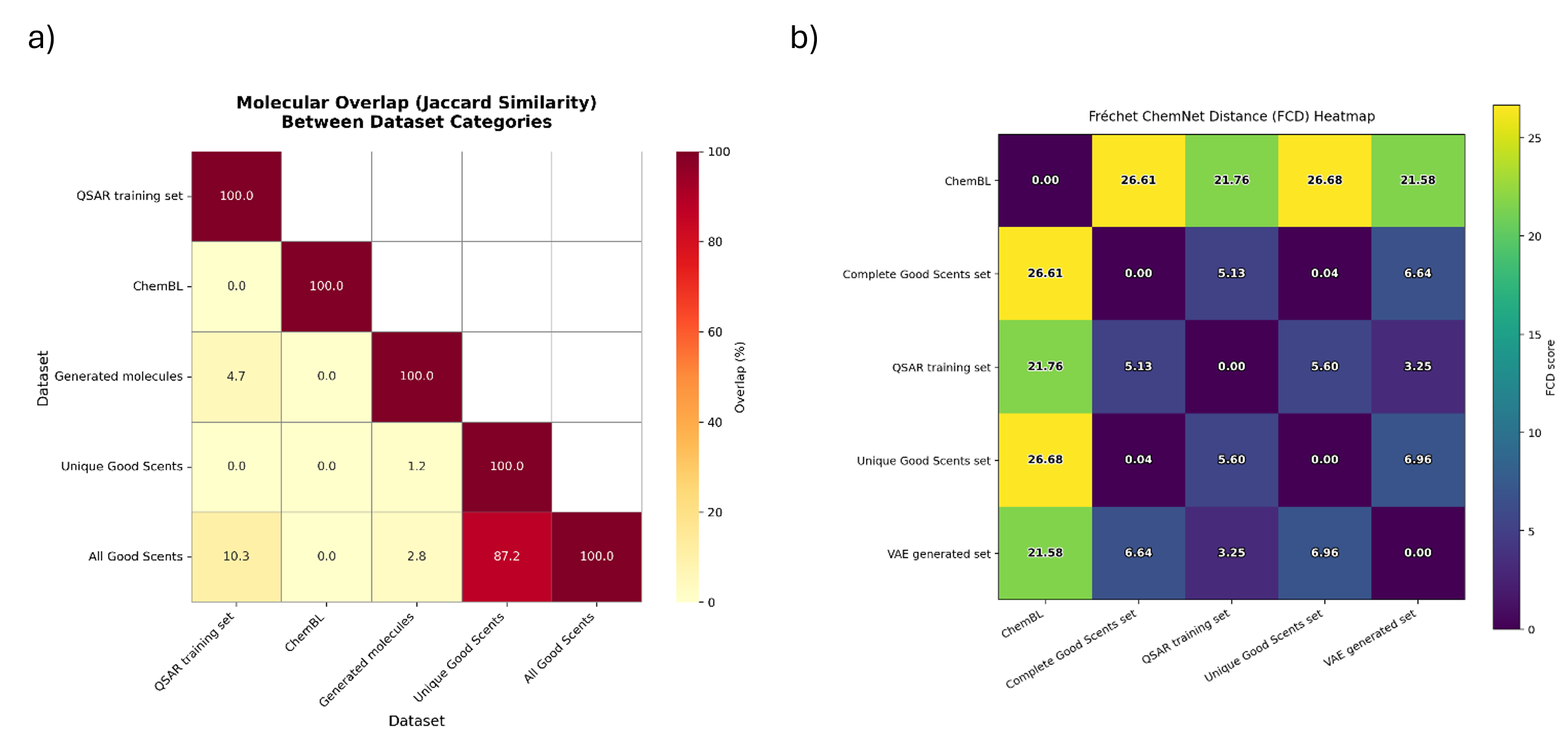}
    \caption{Generated molecules' uniqueness. a) Jaccard similarity, and b) Fréchet ChemNet distance across data-sets.}
    \label{fig:fig5}
\end{figure}

\subsection*{Generated molecule uniqueness}

As a check of generated molecule uniqueness we calculated the Jaccard similarity of the generated odor set against the QSAR model training set, ChEMBL VAE training set and external Good Scents catalog of known  odorants (Figure 5a). 

For validation purposes the Good Scents catalog was split into unique molecules that were unseen during both the QSAR model training and VAE ChEMBL training phases. This `Unique Good Scents' set serves as the primary external ground truth to validate that the QSAR-guided VAE has generalized to relevant chemical space beyond the limitations of the initial logistic regression teacher model. The generated set was found to contain no molecules from ChEMBL and relatively few of the QSAR training set (4.7\% Jaccard similarity). We also found a relatively low Jaccard similarity between the generated set and unique Good Scents molecules (1.2\%), showing that the VAE was capable of discovering many new chemical structures without having seen them previously.

To evaluate the distributional fidelity of the generated chemical space, we calculated the Fréchet ChemNet Distance (FCD), a metric that quantifies the discrepancy between two molecular distributions based on internal activations of the ChemNet neural network [23]. Lower FCD scores indicate greater similarity in chemically and biologically relevant feature space. As shown in Figure 5b, the generated molecules exhibit a striking structural alignment with the known odorant manifold rather than the general chemical background they were trained on. Specifically, the generated set shows a high divergence from the ChEMBL baseline (FCD $\approx$ 21.6) but maintains a close proximity to the QSAR training set (FCD $\approx$ 3.25). Notably, the distributional distance between the generated molecules and the external `Unique Good Scents' catalog (FCD $\approx$ 6.96) is comparable to the intrinsic variance observed between the two ground-truth odor datasets (Training vs. Unique Good Scents, FCD $\approx$ 5.60). This indicates that the QSAR-guided loss shifted the VAE's generative distribution
away from the generic ChEMBL background and toward that of verified odorants; the ablation control
below establishes more precisely what the property head contributes to this
effect.

\begin{table}[ht]
\centering
\caption[Bemis-Murcko Scaffold Classification of Generated Molecules.]{Bemis-Murcko Scaffold Classification of Generated Molecules. The analysis distinguishes between memorization, derivatization of known structures, and the discovery of novel core frameworks. Average properties (Molecular Weight, LogP, and $\log_{10}$ Vapor Pressure) indicate that even novel scaffolds maintain physicochemical profiles consistent with volatility. Averages are arithmetic means over the molecules of each category; $\log_{10}$(VP) is the mean of per-molecule $\log_{10}(P_{\text{sat}}/\text{Pa})$ values, averaged over those molecules carrying a vapor pressure prediction ($n=325$ of $423$ for category 1; complete coverage for categories 2 to 5). Classification was performed on the ring-containing subset of the canonically de-duplicated generated molecules with $P(\text{qsar})>0.99$ ($n=6{,}765$ of $18{,}992$; the remaining $12{,}227$ are acyclic and have no defined scaffold, see Supplementary Table~S1). Category 1 is a molecule-level match; categories 2 to 4 are framework-based and are therefore defined only for ring-containing molecules. Rows 2 to 4 are unaffected by the ring restriction, having no acyclic members. ``In AD'' is the percentage of each category's molecules lying inside the applicability domain, that is, within a nearest-neighbor Tanimoto similarity of $0.35$ on ECFP4 fingerprints to the QSAR training set. The applicability-domain column is reported, not applied: Table~\ref{tab:scaffold_analysis} classifies the full ring-containing library, while the $0.35$ threshold is applied only to the proposed candidate set (Methods, ``Applicability domain''; Results, ``Applicability domain of the QSAR model'').}
\label{tab:scaffold_analysis}
\resizebox{\textwidth}{!}{%
\begin{tabular}{l c c c c c c l}
\toprule
\textbf{Novelty Category} & \textbf{Count} & \textbf{\% of ring-containing} & \textbf{MW} & \textbf{LogP} & \textbf{log$_{10}$(VP)} & \textbf{In AD (\%)} & \textbf{Interpretation} \\
\midrule
1. Exact Memorization & 423 & 6.3 & 157.64 & 2.44 & 1.58 & 100.0 & Model memorized training data (Overfitting). \\
\addlinespace
2. Odorant Derivatization & \multirow{2}{*}{3291} & \multirow{2}{*}{48.6} & \multirow{2}{*}{181.18} & \multirow{2}{*}{2.47} & \multirow{2}{*}{1.18} & \multirow{2}{*}{81.2} & Optimizing side-chains of known \\
(QSAR Scaffold) & & & & & & & odorant frameworks. \\
\addlinespace
3. Repurposing & \multirow{2}{*}{258} & \multirow{2}{*}{3.8} & \multirow{2}{*}{158.31} & \multirow{2}{*}{2.23} & \multirow{2}{*}{1.70} & \multirow{2}{*}{50.8} & Converting generic bioactive \\
(ChEMBL Scaffold) & & & & & & & frameworks into odorants. \\
\addlinespace
4. Validated Scaffold Hop & \multirow{2}{*}{294} & \multirow{2}{*}{4.3} & \multirow{2}{*}{153.27} & \multirow{2}{*}{2.01} & \multirow{2}{*}{2.61} & \multirow{2}{*}{33.0} & True Generalization: Found known odorant \\
(Rediscovery) & & & & & & & frameworks not in training. \\
\addlinespace
5. Uncharted Scaffold Hop & \multirow{2}{*}{2499} & \multirow{2}{*}{36.9} & \multirow{2}{*}{168.95} & \multirow{2}{*}{1.84} & \multirow{2}{*}{2.11} & \multirow{2}{*}{35.4} & Novel frameworks. Potential \\
(New IP) & & & & & & & new chemical entities. \\
\bottomrule
\end{tabular}%
}
\end{table}

\subsection*{Scaffold analysis and novelty depth}
The generated library is predominantly acyclic: 12,227 of the 18,992 candidates at
$P(\text{qsar})>0.99$ (64.4\%) contain no ring system and therefore have no defined
Bemis-Murcko scaffold. The classification below is accordingly reported over the 6,765
ring-containing candidates, with the acyclic majority assessed separately by nearest-neighbor
similarity (below). A predominance of acyclic structures is expected of a generator that has
learned the odorant distribution: the QSAR training set is itself 53.7\% acyclic, and odorant
chemistry is dominated by acyclic esters, aldehydes, alcohols and terpenoid fragments,
consistent with the 0.41 rings per molecule reported below.

To rigorously assess whether the generated molecules represent true structural novelty or merely trivial derivatization (e.g., adding minor functional groups to known structures), we classified the generated set based on Bemis-Murcko scaffolds (Table \ref{tab:scaffold_analysis}). This method abstracts molecules to their core ring-linker frameworks, ignoring side chains.

The analysis reveals that only 6.3\% of ring-containing generated molecules were exact matches to training data, while 48.6\% represented derivatizations of known odorant scaffolds, effectively performing ``lead optimization'' by modifying side chains of validated olfactory cores.

A further 36.9\% of the ring-containing subset falls into the ``Uncharted Scaffold Hop'' category, possessing core frameworks entirely absent from the training sets. The $2{,}499$ molecules in this category span $1{,}429$ distinct Bemis-Murcko frameworks, against $57$ frameworks for the $294$ validated scaffold hops. No framework of this category occurs in any other, so these $1{,}429$ together with the $237$ frameworks of categories 1 to 4 account for all $1{,}666$ distinct frameworks of the ring-containing subset (Supplementary Table~S6). These novel scaffolds exhibit physicochemical properties ideal for volatiles (Mean MW $\approx$ 170~Da, $\log_{10}(\text{VP}) \approx 2.11$), suggesting they represent plausible new chemical entities rather than generation artifacts. Furthermore, the model achieved a ``Validated Scaffold Hop'' rate of 4.3\%, effectively rediscovering distinct odorant frameworks found in the external ``Unique Good Scents'' validation set that were never presented during training. This indicates the model has an ability to generalize structural rules of olfaction to discover valid, novel molecular architectures.

Because Bemis-Murcko analysis is undefined for acyclic compounds, structural novelty
across the whole library was additionally assessed by nearest-neighbor Tanimoto similarity to
the union of the reference sets. The acyclic subset is comparably distant from known chemistry to
the ring-containing subset (mean nearest-neighbor similarity 0.486 against 0.476; 35.4\%
against 40.5\% below 0.4; Supplementary Table~S4, assessed over the full library), so the
majority of the library that
Bemis-Murcko cannot describe does not consist of trivial variants of known compounds. This
scaffold-independent estimate agrees closely with the scaffold-hop rate above. The conventional
per-scaffold measure of diversity is reported with its control in Supplementary
Table~S5.

\begin{figure}[tp]
  \centering
  \includegraphics[width=0.85\textwidth]{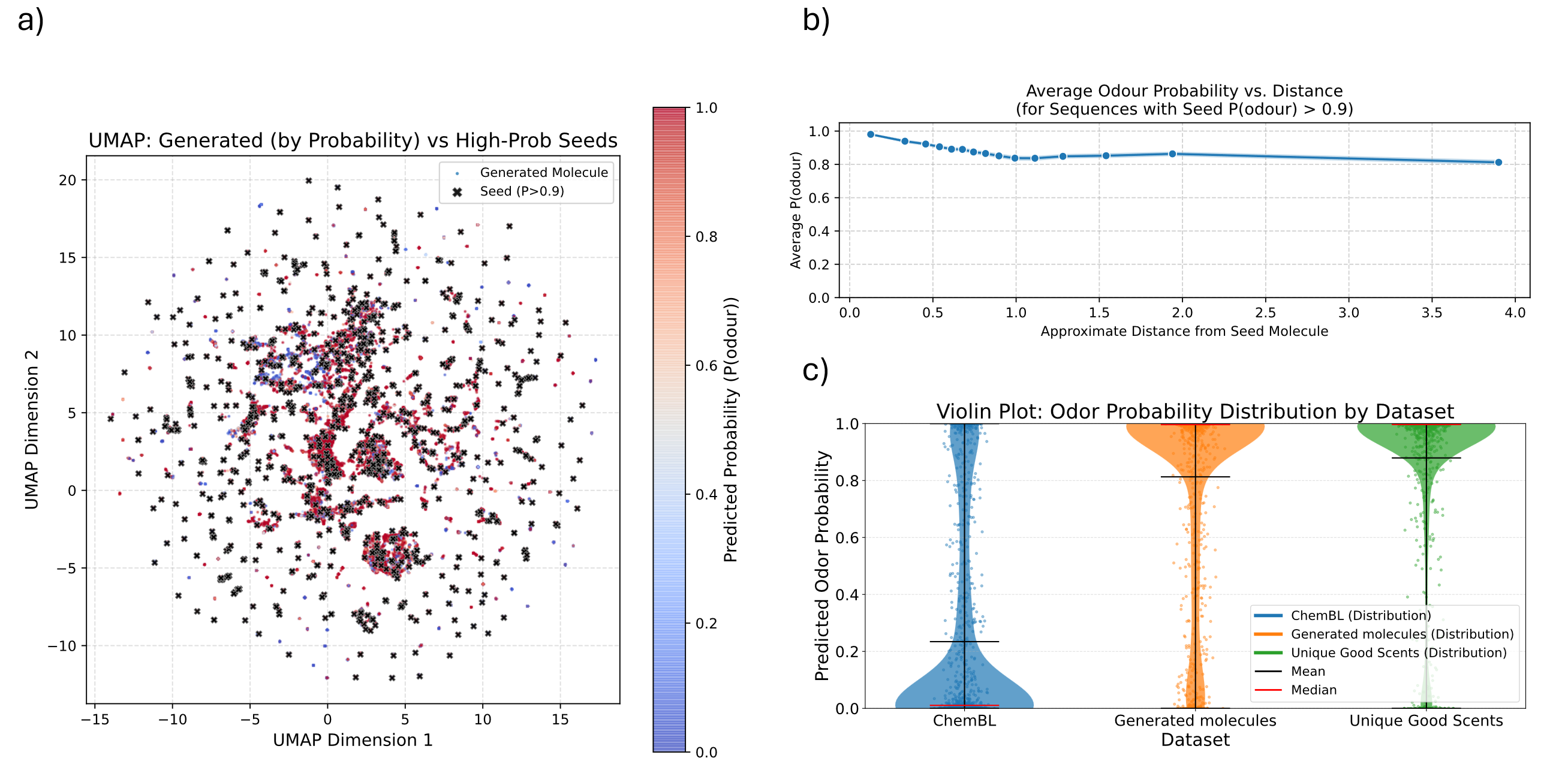}
  \caption{Generated molecules odor probability. a) UMAP representation of seed and generated molecules encoded in the latent space color-coded by odor (inset shows the $P(\text{qsar})$ distribution across all generated molecules). b) probability of generated molecule being odorous with distance from the seed (in units of $\sigma_{\phi}(z|X)$)  averaged over all generated molecules (thickness shows 95\% confidence interval). c) QSAR model predicted odor probabilities of random 1k molecules sampled from ChEMBL, generated molecules set and unique Good Scents (ground truth compounds).}
  \label{fig:fig6}
\end{figure}

\subsection*{Applicability domain of the QSAR model}
Predictions from any QSAR model are reliable only within the region of chemical space the model
was trained on, and a generative framework is by construction liable to produce structures
outside it. We therefore quantified domain membership across the generated library. Of the
18,992 candidates at $P(\text{qsar})>0.99$, 93.0\% fall inside the domain at the training-derived
threshold of $0.2465$, 68.5\% at $0.35$, and 55.7\% at $0.40$.

The predicted odor probability is, however, almost uninformative about domain membership. Mean
$P(\text{odor})$ is $0.818$ for candidates inside the domain against $0.777$ for those outside,
a difference of $0.04$ to $0.05$ that persists at every threshold examined. A high predicted
probability therefore does not indicate that the prediction is reliable, and domain membership
must be applied as a separate criterion rather than inferred from model confidence.

Domain coverage declines monotonically with structural novelty. Among ring-containing candidates
at the $0.35$ threshold, exact memorizations are retained in full (100\%), derivatizations of
known odorant scaffolds at 81.2\%, repurposed ChEMBL scaffolds at 50.8\%, uncharted scaffold
hops at 35.4\% and validated scaffold hops at 33.0\%. The mechanism is structural rather than
incidental: compounds in the last two categories are defined by bearing scaffolds absent from
the QSAR training set, so a domain defined by proximity to that set must necessarily under-cover
them. This is a general limitation of applicability-domain filtering in generative work, and it
implies that such filtering preferentially retains the least novel part of a generated library.
We therefore report domain membership alongside the classification (Table~\ref{tab:scaffold_analysis})
rather than applying it as a filter to the analysis itself.

Domain membership and structural novelty are nonetheless distinct properties rather than
opposite ends of a single axis, because they are measured against different references: domain
membership against the QSAR training set, which defines what the predictor has seen, and novelty
against ChEMBL and the Unique Good Scents catalog, which define what is structurally known.
Scored this way, 4,844 of the 18,992 candidates (25.5\%) lie inside the
applicability domain at the $0.35$ threshold while remaining below $0.4$ nearest-neighbor
similarity to both non-training references. The proportion is near-identical for the two subsets,
26.5\% of the acyclic candidates and 23.7\% of the ring-containing ones, so this is a property of
the library rather than of a convenient fraction of it. It does not make predictions issued far
outside the domain reliable; what it shows is that the region which is novel relative to known
odorant chemistry is not the same region as that which is unfamiliar to the predictor.

\paragraph{Candidate selection:}
For the purpose of proposing candidates for experimental
evaluation, rather than characterizing the generated library, two filters are applied in
sequence. Of the 18,992 candidates at $P(\text{qsar})>0.99$, 13,012 (68.5\%) fall inside the
applicability domain at the $0.35$ threshold, and 12,720 remain after removal of reactive
chemotypes (Methods, ``Reactive chemotype screening''; Results, ``Generated molecule safety''). The two filters are independent and neither
substitutes for the other: 81.1\% of the reactive compounds identified lie inside the
applicability domain, since the domain measures where the predictor is reliable and carries no
information about whether a compound is safe to inhale.

\subsection*{Latent space organization and structure-odor relationships}
Generated compounds were found to vary continuously in odor likelihood with distance from a high probability odorant seed in the latent space (Figure 6a,b); however, sometimes small changes in the molecular structure could also dramatically change $P(\text{odor})$ (generative examples within Supplementary Data). Such discontinuities highlight the non-linear mapping between the continuous latent representation of molecular structure and the predicted biological property. This observation aligns with the concept of activity cliffs in structure-odor relationships (SORs), where minor chemical alterations can lead to significant perceptual shifts [24].

The latent space retains the smooth `chemical grammar' necessary for generating valid molecules via interpolation (Figure 4 and Supplementary Data examples), while the integrated odor prediction capability reflects the sensitivity of olfactory perception to specific structural features.
Running the generated molecule set through the QSAR model demonstrates that the distribution is distinct from the ChEMBL grammar learning data-set whilst closer to the Unique Good Scents ground truth set (Figure 6c).

The finding that predicted odor probability drops only modestly on average (remaining > 0.8 even at
significant distances from seed molecules) shows that the highest concentration of generated
odors are within a close radius of a seed's latent representation (Figure 6b), the persistence of
elevated probability further out indicates that regions corresponding to odorous
molecules are not isolated points but likely form continuous, potentially interconnected
neighborhoods within the latent space (Figure 6b shows a small
increase in probability at the distance from the seed molecule close to that of neighboring seeds
on average). As $P(\text{odor})$ remains relatively high over distances comparable to the average
separation between known distinct odorants it reinforces the idea that the model can effectively
bridge different regions of odor space. As established by the ablation control below,
this behavior is shared by a VAE trained without the property head and does not by itself
demonstrate that the odor prediction loss organized the latent space; what the loss contributes is
the density of retrievable structures within that neighbourhood. Practically, it means that sampling around known odorants is
an effective strategy for discovering novel candidates that retain desirable olfactory properties,
confirming the potential for guided exploration within the chemical space.

By adjusting the loss function coefficients ($\beta_{KL}, \gamma$ in Equation 12) the framework
allows for adjusting the trade-off between maintaining strict chemical continuity/similarity
(grammar) and maximizing the desired odor property. This adjustability enables researchers to
either perform broad exploration of chemically plausible space or focus generation efforts
(`exploitation') specifically towards regions predicted to yield high odor probability, depending on the goals.

\subsection*{Ablation control: the odor prediction head structures the latent space}

An alternative explanation must be excluded for two of the results reported above: the
high $P(\text{qsar})$ retained by compounds decoded near an odorant seed (Figure~\ref{fig:fig6}b),
and the alignment of the generated distribution with known odorants rather than with ChEMBL
(Figure~\ref{fig:fig5}b). Molecules decoded from the neighbourhood of a seed are structurally
similar to it and, by the similarity-property principle, may inherit a baseline of odor-likeness
\emph{irrespective of the training objective}; a generic VAE, never told anything about odor, might
therefore reproduce both observations. To separate the two explanations we repeated the
seeded-generation analysis on the ablated model defined in Methods, identical to the full model but
for removal of the property loss.

\paragraph{Protocol:}
Both models were seeded with the same $1{,}615$ odorant compounds and sampled with the
protocol described above ($N=100$ decode attempts per seed), using the $1{,}142$ seeds that both
models completed, restricted to the $984$ of those with $P(\text{qsar})>0.9$ and to \emph{novel}
decodes, i.e.\ excluding the seed returned unchanged. Two controls are essential. First, both sets
were scored with the \emph{same} external QSAR model: the ablated model has no property head and
emits no $P(\text{odor})$, and a single independent probe is in any case the correct design, since
it measures the chemistry each latent space yields rather than each model's opinion of its own
output. Second, distance was normalized \emph{per model}, as the two latent spaces are not on a
common scale --- the mean pairwise inter-seed distance is $2.59$ in the full model against $1.57$
in the ablated model, a $1.65\times$ contraction --- so raw distances would compress the ablated
abscissa and manufacture an apparent difference in fall-off.

\begin{figure}[htbp]
  \centering
  \includegraphics[width=0.92\textwidth]{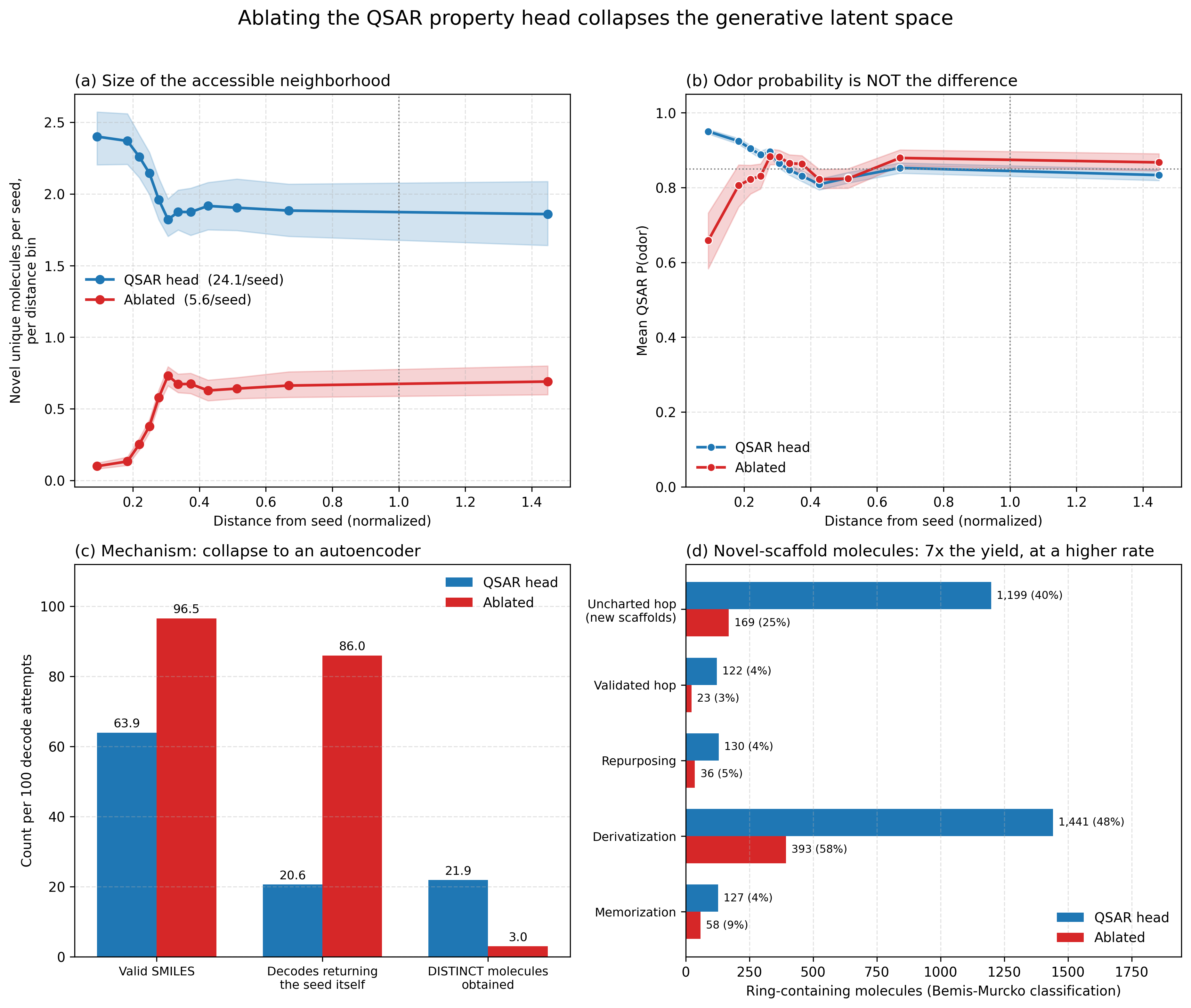}
  \caption{Ablation of the odor prediction head. Both models share an identical
architecture, training corpus and training schedule, differing only in the presence of the QSAR
property loss ($\gamma\mathcal{L}_{P(\text{odor})}$, Equation~\ref{eq:12}); both were seeded with
the same $984$ odorant compounds ($P(\text{qsar})>0.9$) and given $100$ decode attempts per seed.
Distance is normalized by each model's own mean pairwise inter-seed distance ($2.59$ with the head,
$1.57$ ablated), so that $1.0$ (dotted vertical line) corresponds to the typical separation between
two seed odorants; without this normalization the $1.65\times$ contraction of the ablated latent
space would itself produce an apparent difference. \textbf{(a)} Distinct novel molecules per seed
as a function of distance; shading gives the 95\% confidence interval bootstrapped over seeds.
\textbf{(b)} Mean QSAR odor probability of the same compounds, both models scored with the same
external logistic-regression model. \textbf{(c)} Zero-perturbation decoding control, each seed's
own latent mean decoded $100$ times: valid strings emitted, decodes returning the seed itself, and
distinct molecules obtained. \textbf{(d)} Bemis-Murcko classification of the novel compounds
($P(\text{qsar})>0.99$) through the pipeline of Table~\ref{tab:scaffold_analysis}, restricted to ring-containing molecules. Values are
given in the text.}
  \label{fig:ablation}
\end{figure}

\begin{table}[H]
\centering
\caption{Bemis-Murcko scaffold classification of the novel compounds generated by the
full model and by the ablated model (no odor prediction head), from an identical seed set ($984$
seeds, $P(\text{qsar})>0.9$) and an identical budget of $100$ decode attempts per seed. Both sets
were classified against the same three reference sets (QSAR training set, ChEMBL, Unique Good
Scents) and filtered at $P(\text{qsar})>0.99$, exactly as in
Table~\ref{tab:scaffold_analysis} and restricted to ring-containing molecules, as defined in Methods (``Scaffold classification'').}
\label{tab:ablation_scaffold}
\begin{tabular}{l r r r r}
\toprule
 & \multicolumn{2}{c}{\textbf{Full model}} & \multicolumn{2}{c}{\textbf{Ablated}} \\
\cmidrule(lr){2-3} \cmidrule(lr){4-5}
\textbf{Novelty category} & \textbf{Count} & \textbf{\% Gen.} & \textbf{Count} & \textbf{\% Gen.} \\
\midrule
1. Exact Memorization                       &     127 &  4.2 &   58 &  8.5 \\
2. Odorant Derivatization (QSAR scaffold)   &   1,441 & 47.7 &   393 & 57.9 \\
3. Repurposing (ChEMBL scaffold)            &     130 &  4.3 &    36 &  5.3 \\
4. Validated Scaffold Hop (rediscovery)     &     122 &  4.0 &    23 &  3.4 \\
5. Uncharted Scaffold Hop (new IP)          &   1,199 & 39.7 & 169 & 24.9 \\
\midrule
\textbf{Total classified}                   &  3,019 &       & 679 &       \\
\bottomrule
\end{tabular}
\end{table}

The outcome is more specific than we had anticipated: the property head does not raise
the odor probability of the molecules that are generated --- it determines how many are generated
at all, and it shifts the composition of what is generated toward novel frameworks.

\paragraph{Odor probability alone does not discriminate the two models:}
Averaged over normalized distance the two models are indistinguishable in odor-likeness
(Figure~\ref{fig:ablation}b): mean $P(\text{qsar})=0.873$ against $0.853$, medians $0.997$ and
$0.995$, and $77.5\%$ against $75.3\%$ of compounds exceeding $P(\text{qsar})>0.9$. Both traces
remain above $\approx\!0.85$ out to and beyond one mean inter-seed distance and their 95\%
confidence intervals overlap across most of the range. This is precisely what the
similarity-property argument predicts, and we accept its consequence: the persistence of high
$P(\text{qsar})$ with distance, reported for the full model in Figure~\ref{fig:fig6}b, is
\emph{not} on its own evidence that the property loss has organized the latent space, and we do not
claim it as such.

\paragraph{The QSAR head determines the size of the accessible neighbourhood, and why:}
The models differ decisively in generative yield (Figure~\ref{fig:ablation}a): from the
same $984$ seeds and the same decode budget the full model returns $23{,}756$ distinct novel
compounds ($24.1$ per seed) against $5{,}507$ ($5.6$ per seed), a $4.3$-fold reduction, with the
separation present at every distance and 95\% confidence intervals --- bootstrapped over seeds, the
independent units of replication --- that do not overlap at any point. The mechanism is exposed by
a control in which no perturbation is applied and each seed's own latent mean is decoded $100$
times (Figure~\ref{fig:ablation}c): the ablated model emits $96.5$ valid SMILES per $100$ attempts
but returns the seed itself on $86.0$ of them, leaving only $3.0$ distinct compounds per seed,
whereas the full model emits $63.9$ valid strings, returns the seed on $20.6$, and yields $21.9$
distinct compounds. On both metrics that could explain a diversity deficit --- syntactic validity
and reconstruction fidelity --- the ablated decoder is substantially \emph{better}, so the yield
gap cannot be attributed to decoder quality. Absent a second objective competing for latent
capacity the ablated model has become a near-deterministic autoencoder, spending that capacity on
reproducing its input; requiring $z$ to carry odor information in addition to structure flattens
the decoder's output distribution, so that repeated sampling at a single latent point explores a
neighbourhood of related odorants rather than re-deriving the seed. This also accounts for the
$1.65\times$ contraction of the ablated latent space.

\paragraph{Scaffold-level exploration separates the two models:}
Classifying the novel compounds of both models through the identical pipeline used for
Table~\ref{tab:scaffold_analysis} (Table~\ref{tab:ablation_scaffold},
Figure~\ref{fig:ablation}d), the rate of scaffold hopping differs by 14.8 percentage points ($39.7\%$ uncharted
hops against $24.9\%$), and the absolute yield by a larger factor still:
$1{,}199$ against $169$ molecules bearing scaffolds absent from every training set, spanning
$764$ distinct novel frameworks against $145$, from an identical seed set and compute
budget. The one rate that shifts materially is exact memorization, which more than
doubles on ablation ($4.2\%\rightarrow8.5\%$), consistent with the collapse described above. The
full model retrained here recovers an uncharted-hop rate of $39.7\%$, in close agreement with the
$36.9\%$ reported in Table~\ref{tab:scaffold_analysis} for the original model --- an independent
replication of that result on a separately trained network.

\paragraph{Interpretation:}
The ablation refines, rather than merely corroborates, our earlier claim: the statement
that the QSAR-guided loss ``steered the VAE's latent space away from the generic ChEMBL
distribution'' should be understood in terms of the \emph{extent and navigability} of the odorous
region rather than the odor-likeness of any single point within it. A conventional VAE already
places structurally similar --- and hence plausibly odorous --- compounds near an odorant seed;
what it does not do is make that region generatively accessible, its latent space being contracted
and its decoder nearly deterministic, so that the neighbourhood, though notionally odorous, is
almost empty of retrievable alternatives. The property head converts a region that merely
\emph{is} odorous into one that can be \emph{mined}, and it is this that underwrites the downstream
results of this paper: the distributional alignment with known odorants (Figure~\ref{fig:fig5}b),
the $36.9\%$ uncharted scaffold-hop rate (Table~\ref{tab:scaffold_analysis}), and the
$10$--$25\times$ multiplier on the seed set, each of which presupposes a generated library large
and diverse enough to support the analysis.

\subsection*{Generated molecule validity checks}
To assess whether the generated SMILES sequences translate into physically realizable 3D entities, we subjected the \textit{de novo} molecules to a battery of stringent validity checks. 
Beyond standard syntactic correctness, which was enforced to 100.00\% across both RDKit and OpenBabel parsers via rejection sampling, the generated candidates demonstrated exceptional geometric and physical plausibility. Notably, 99.90\% of generated structures complied with Bredt’s rule regarding double bond placement at bridgeheads—surpassing the compliance rates of both the training set (99.27\%) and the general ChEMBL baseline (98.27\%).
This structural integrity is further supported by energy strain calculations, where 99.38\% of generated conformers fell within acceptable energy limits, aligning closely with the `Unique Good Scents' ground truth. Furthermore, the model balanced structural novelty with medicinal chemistry viability; while 93.94\% of the generated molecules were unique (novel entities), they maintained a low PAINS alert rate (2.26\%) comparable to the QSAR training set (2.49\%), indicating that the generative process explores new chemical space without introducing excessive structural liabilities or frequent assay interference artifacts.

\begin{table}[h]
\centering
\caption{Combined Validity, Energy, Structural, and Novelty Metric Checks}
\label{tab:combined_metrics_full}
\resizebox{\textwidth}{!}{%
\begin{tabular}{l ccc c c cc cc}
\toprule
 & \multicolumn{3}{c}{\textbf{Chemical Validity Checks}} & \textbf{Geometric} & \textbf{Physical} & \multicolumn{2}{c}{\textbf{Structural Alerts (PAINS)}} & \multicolumn{2}{c}{\textbf{Novelty / Overlap}} \\
\cmidrule(lr){2-4} \cmidrule(lr){5-5} \cmidrule(lr){6-6} \cmidrule(lr){7-8} \cmidrule(lr){9-10}
\textbf{Dataset} & \textbf{RDKit} & \textbf{OpenBabel} & \textbf{Cross} & \textbf{Bredt} & \textbf{Energy Strain} & \textbf{Clean /} & \textbf{PAINS} & \textbf{Unique} & \textbf{Overlap} \\
 & \textbf{Valid (\%)} & \textbf{Valid (\%)} & \textbf{Valid (\%)} & \textbf{Compliant (\%)} & \textbf{Acceptable (\%)} & \textbf{Free (\%)} & \textbf{Alerts (\%)} & \textbf{(\%)} & \textbf{(\%)} \\
\midrule
Generated molecules & 100.00 & 100.00 & 100.00 & 99.90 & 99.38 & 97.74 & 2.26 & 93.94 & 6.06 \\
QSAR Training Set & 100.00 & 100.00 & 100.00 & 99.27 & 97.00 & 97.51 & 2.49 & 95.74 & 4.26 \\
Unique Good Scents & 100.00 & 100.00 & 100.00 & 96.85 & 99.60 & 99.24 & 0.76 & 99.08 & 0.92 \\
All Good Scents & - & - & - & 97.13 & 99.40 & 99.22 & 0.78 & 97.45 & 2.55 \\
ChEMBL & 100.00 & 100.00 & 100.00 & 98.27 & 93.45 & 98.65 & 1.35 & 100.00 & 0.00 \\
\bottomrule
\end{tabular}%
}
\end{table}

\begin{figure}[tp]
  \centering
  \includegraphics[width=1\textwidth]{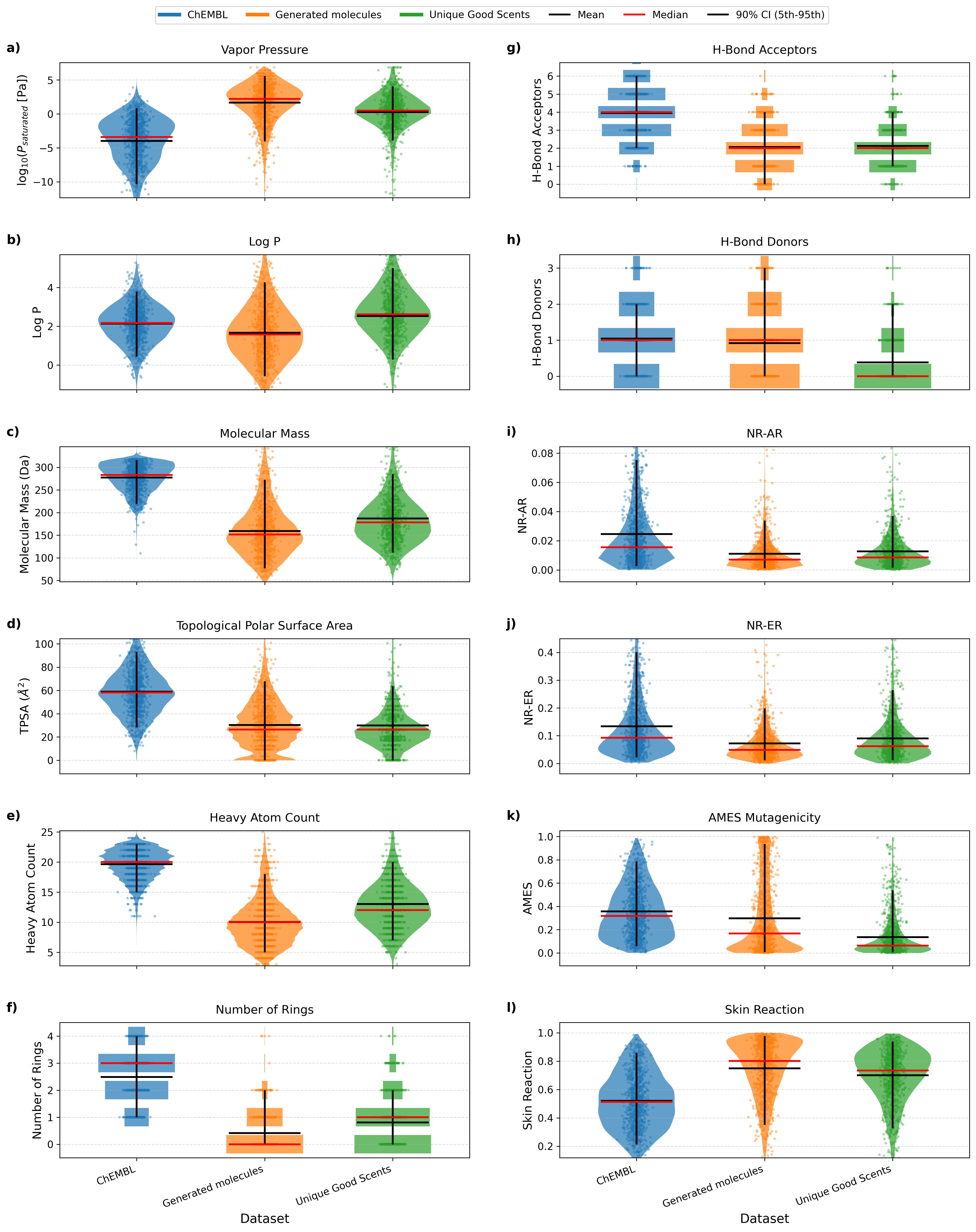}
    \caption{Physicochemical and ADMET property validation on generated molecules compared against ChEMBL training molecules and unique Good Scents (ground truth compounds) . a-h) Predicted physicochemical properties, i-l) Predicted ADMET properties, NR-AR: Nuclear Receptor – Androgen Receptor; NR-ER: Nuclear Receptor – Estrogen Receptor;  AMES - Ames Mutagenicity Test; Skin Reaction. Panels f--h show integer-valued counts as discrete bars giving the relative frequency of each value; the remaining panels show kernel-density violins.}
    \label{fig:fig7}
\end{figure}

\subsection*{Generated molecule physicochemical properties}
The generated molecules exhibit physicochemical profiles that align closely with the established criteria for odorants, distinct from the broader chemical space represented by ChEMBL (Figure \ref{fig:fig7}). This comparison against the ChEMBL baseline serves to validate that the model successfully targets the specific physicochemical niche of odorants, confirming the chemical relevance of the generated entities irrespective of the specific generative architecture employed. A critical determinant of olfactory perception is volatility; the generated molecules display a mean $\log_{10}(P_{\text{sat}}/\text{Pa})$ of 1.66 (calculated using \textit{VaPoRs} [25]), which is significantly higher than the non-odorant ChEMBL baseline (-3.97 Pa) and comparable to the Unique Good Scents ground truth (0.26 Pa), ensuring sufficient volatility to reach the nasal epithelium [5]. Molecular weight analysis further corroborates this trend, with the generated set having a mean mass of 159.29 Da, effectively mirroring the lighter distribution of known odorants (186.79 Da) compared to the heavier ChEMBL average (277.09 Da). This lower molecular mass is consistent with the requirement for rapid diffusion and receptor interaction [5] and the QSAR training set properties (generated molecules set mean MW$=$$159.29 \pm 57.07$ Da vs. QSAR training set MW$=$$153.32 \pm 71.86$ Da).  In terms of hydrophobicity, the generated molecules possess a mean Log P of 1.67, falling well within the "Goldilocks" zone (typically 1--3) required for bioavailability and crossing the mucosal layer, whereas the ChEMBL set extends into more lipophilic regions (mean 2.14). Structural complexity, measured by the number of rings and heteroatoms, also reflects the ground truth; 
generated molecules average 0.41 rings and 2.07 hydrogen bond acceptors, tracking the Good Scents distribution (0.81 rings, 2.13 acceptors) far more closely than the ChEMBL dataset (2.48 rings, 3.93 acceptors). The Kullback-Leibler divergence scores confirm this structural fidelity, showing minimal divergence between generated and ground truth distributions (e.g., Log P KL $\approx$ 0.18) compared to the significant separation from ChEMBL (Log P KL $\approx$ 0.40).

\subsection*{Generated molecule safety}
To assess the safety and suitability of the generated compounds for potential application, ADMET properties were predicted using ADMET-AI [26]. The generated set demonstrates a favorable predicted profile on the endpoints assessed, with a low predicted probability of nuclear receptor toxicity. Specifically, the mean predicted activity for the Androgen Receptor (NR-AR) was 0.01, and for the Estrogen Receptor (NR-ER) was 0.07, values that are statistically indistinguishable from the Unique Good Scents catalog (NR-AR: 0.01; NR-ER: 0.09) and significantly lower than the ChEMBL baseline (NR-AR: 0.02; NR-ER: 0.13). Mutagenicity potential, assessed via predicted Ames test outcomes, showed a mean probability of 0.30 for the generated molecules. While this is slightly elevated compared to the ground truth (0.14), it remains lower than the ChEMBL background (0.36), suggesting that the generative process does not disproportionately introduce genotoxic structural alerts. Interestingly, the predicted probability of skin sensitization was higher in both the generated set (0.75) and the Good Scents collection (0.70) compared to ChEMBL (0.52). This elevation likely reflects the reactive nature of many volatile organic compounds (such as aldehydes and enones) common in fragrance chemistry, highlighting a known trade-off between olfactory potency and potential dermal reactivity that requires standard regulatory management rather than indicating a failure of the generative model.

Predicted ADMET profiles do not identify compounds carrying reactive functional groups,
which may be detectable by smell while being irritant rather than odorant. Structural screening
(Methods, ``Reactive chemotype screening'') identified 360 such compounds, 1.9\% of the 18,992 candidates at $P(\text{qsar})>0.99$,
against 3.5\% in the QSAR training set and 1 of 4{,}343 in the Unique Good Scents catalog.\footnote{The single match in the Unique Good Scents catalog is a 587~Da
rosin-derived diterpene anhydride with woody and balsamic odor descriptors. The pattern
matches carboxylic anhydrides irrespective of size or reactivity; the screen identifies
chemotypes for review rather than adjudicating hazard.} These are
excluded from the proposed candidate set. Of the compounds identified, 81.1\% lie inside the
applicability domain, so domain membership provides no protection against this class: the two
screens are independent.

\subsection*{Molecule synthesis accessibility}
\begin{figure}[t!]
  \centering
  \includegraphics[width=1\textwidth]{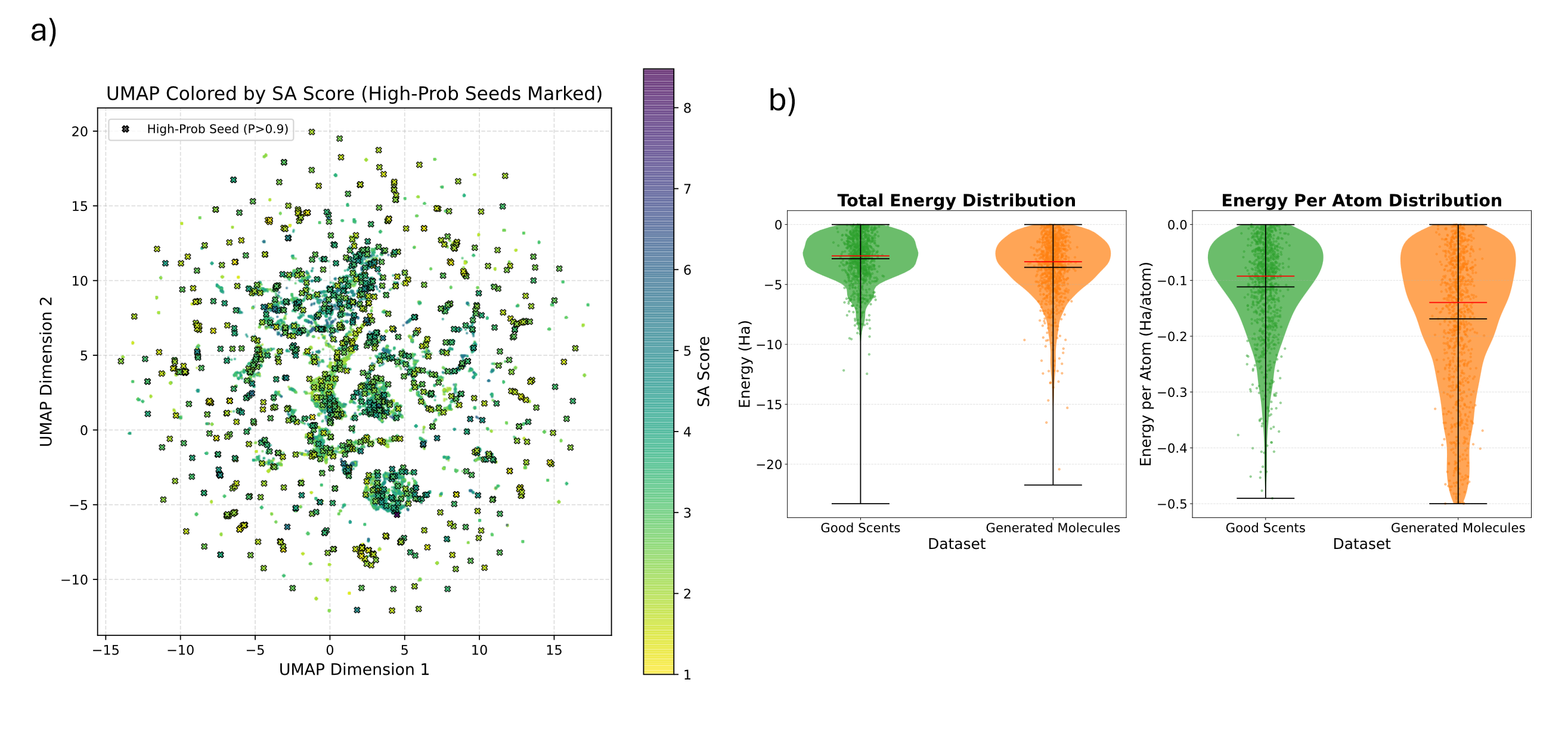}
    \caption{Synthesis accessibility and quantum chemistry checks. a) UMAP representation of seed and generated molecules encoded in the latent space color-coded by synthesis accessibility (SA – lower score relates to simpler synthesis complexity), b) comparison of total molecule and per atom energy calculations between generated molecules and Good Scents (ground truth molecules).}
    \label{fig:fig8}
\end{figure}

We also compared the synthesis accessibility (SA) for generated molecules by computing the
SAScore [27] using RDKit's SAScorer (Figure \ref{fig:fig8}a). This score is based on fragment contributions and includes a
penalty for molecular complexity. While SA scores for generated compounds were found to be
comparable to seed molecules, generated compounds have a slightly higher mean SA score of 3.12 ± 1.00 compared with seed molecules with mean SA score 2.51 ± 0.97, possibly on account of increased diversity in the generated set and the fact that seed compounds are often optimized for their synthesis routes in the literature and public data-sets.
\begin{table}[h!]
\centering
\resizebox{\textwidth}{!}{%
\begin{tabular}{lrrrrrrrrrrrr}
\toprule
\textbf{Data-set} & \textbf{\% Solved} & \textbf{Avg state} & \textbf{Min state} & \textbf{Max state} & \textbf{State score} & \textbf{Avg synth} & \textbf{Min synth} & \textbf{Max synth} & \textbf{Synth Routes} & \textbf{Avg} & \textbf{Avg} & \textbf{Avg precurs.} \\
 & & score & score & Score & st. dev. & routes & routes & routes & st. dev & reactions & precursors & in stock \\
\midrule
\textbf{Generated molecules} & 100.00 & 0.76 & 0.05 & 1.00 & 0.27 & 5.30 & 1.00 & 25.00 & 2.71 & 2.89 & 2.54 & 2.11 \\
Unique Good Scents & 100.00 & 0.92 & 0.05 & 1.00 & 0.17 & 4.10 & 1.00 & 16.00 & 2.31 & 1.85 & 1.99 & 1.71 \\
ChEMBL & 100.00 & 0.90 & 0.05 & 1.00 & 0.15 & 4.85 & 1.00 & 21.00 & 2.50 & 2.65 & 2.67 & 1.71 \\
All Good Scents & 100.00 & 0.93 & 0.05 & 1.00 & 0.17 & 3.82 & 1.00 & 15.00 & 2.15 & 1.61 & 1.83 & 1.60 \\
\bottomrule
\end{tabular}%
} 
\caption{Retrosynthesis analysis of data-sets}
\label{tab:retrosynthesis}
\end{table}

To move beyond theoretical validities and assess practical constructability, we performed a comprehensive retrosynthetic analysis using AiZynthFinder [28], a template-based Monte Carlo Tree Search (MCTS) algorithm trained on the USPTO reaction database. As detailed in Table \ref{tab:retrosynthesis}, the generated molecules showed high predicted synthetic accessibility, with the algorithm identifying valid synthesis routes for 100.00\% of the candidates within the specified search limits. This 100\% solvability rate matches the ground-truth ``Unique Good Scents'' and ``All Good Scents'' datasets, indicating that the generative model implicitly learned chemically stable motifs amenable to standard organic transformations. Crucially, the ``State Score''—a machine-learning-derived metric representing the confidence and feasibility of the identified routes (where 1.0 represents high confidence)—averaged 0.76 for the generated set. While slightly lower than the highly optimized commercial odorants in the Good Scents database (0.92--0.93), a score of 0.76 remains well within the range of experimentally actionable chemistry. The standard deviation of 0.27 in the generated set, compared to the tighter distributions of the training sets (0.15--0.17), reflects the model's exploration of novel chemical space. The high solvability and low average step count ($\sim$2.89 steps) are likely influenced by the lower average molecular weight of the generated library, as smaller volatile molecules are generally more accessible from commercial precursors. This suggests the VAE is not merely memorizing easy-to-make training examples but is venturing into novel structural territories while retaining synthetic accessibility.

The practical viability of the generated candidates is further supported by route complexity metrics and precursor availability from the ZINC stock database. The generated molecules required, on average, 2.89 reaction steps to synthesize, a value that sits between the simple commercial odorants (1.61--1.85 steps) and the more complex general bioactive compounds in ChEMBL (2.65 steps). This average step count of $\sim$3 is ideal for industrial R\&D, balancing structural novelty with process efficiency—a critical factor given that odorants are often low-molecular-weight ($<300$ Da) compounds where long synthetic sequences are economically prohibitive. Furthermore, the analysis identified an average of 2.11 commercially available precursors in stock per generated molecule, surpassing the availability metrics for both the unique Good Scents (1.71) and ChEMBL (1.71) datasets. This high precursor availability (average stock availability ratio of $>70$\% per route) significantly de-risks downstream experimental validation, as the majority of starting materials can be sourced directly rather than synthesized \textit{de novo}. The finding that the generated set maintains short reaction pathways (max 25.00 routes found) using readily available building blocks is consistent with the QSAR-guided latent space encoding not just bioactivity and validity, but also the pragmatic constraints of synthetic organic chemistry.

\subsection*{Quantum chemical stability analysis}
To validate the thermodynamic stability of the generated structures beyond simple geometric rules, we performed semi-empirical quantum mechanical calculations using the GFN2-xTB method [29]. This approach provides a rigorous assessment of the electronic ground state energy, offering a proxy for chemical stability. As shown in Figure \ref{fig:fig8}b, the total energy distribution of the generated molecules aligns remarkably well with the ground-truth ``Unique Good Scents'' dataset. The generated molecules exhibited a mean total energy of $-3.57$ Hartree (Ha), compared to $-2.85$ Ha for the known odorants, with the slightly lower (more stable) energy in the generated set largely attributable to the minor molecular weight differences discussed previously. When normalized for molecular size, the energy per atom distributions are also statistically convergent. The generated molecules possess a mean energy per atom of $-0.17 \pm 0.12$ Ha, which overlaps significantly with the Good Scents distribution ($-0.11 \pm 0.08$ Ha). The low Kullback-Leibler divergence between the two distributions (KL $\approx$ 0.14) confirms that the generative model does not produce high-energy, strained conformers or unstable electronic configurations. Instead, it successfully targets the thermodynamically favorable region of chemical space characteristic of stable, volatile organic compounds found in fragrance palettes.

\subsection*{Comparative studies}
Quantitative benchmarking against emerging generative frameworks reveals that the proposed QSAR-guided VAE offers a distinct balance of high novelty and industrial readiness. While Rodrigues et al. (2024) [30] utilize GNNs primarily for reformulation—explicitly excluding novel compounds to ensure database matches—and Sharma et al. (2025) [31] provide a valuable architectural benchmark of Diffusion and Transformer models, our framework is engineered specifically for \textit{de novo} discovery. In terms of generative performance, our model matches the perfect validity of Sharma’s Diffusion model ($100\%$) but significantly outperforms it in structural novelty ($93.9\%$ vs. $87.0\%$). Crucially, where Sharma’s Diffusion model exhibits high scaffold similarity ($0.64$), indicating a tendency toward derivatization, our VAE achieves a $36.9\%$ rate of uncharted scaffold hops, discovering fundamentally new odorant cores. Additionally, our targeted QSAR model demonstrates superior discrimination (F1 score: $0.97$) compared to the best-performing logistic benchmark in Sharma et al. (F1 score: $0.94$). Finally, we address the ``realism gap'' identified as a limitation in these concurrent studies; by extending validation beyond heuristic rules to include $100\%$ predicted synthetic solvability (AiZynthFinder), ADMET profiling, and thermodynamic stability (GFN2-xTB), we provide the actionable downstream data required to transition generated candidates from theoretical algorithms to physical chemical development.

Rodrigues et al. (2024) employs Graph Neural Networks (GGNNs) for fragrance optimization but explicitly excludes novel chemical entities from the final selection, relying instead on ``rediscovering'' known catalog compounds to ensure odor validity. In stark contrast, our framework does not merely retrieve knowns or generate theoretical structures; it actively discovers \textit{de novo} entities, achieving a $36.9\%$ rate of uncharted scaffold hops. Critically, we bridge the gap between generation and physical realization where other studies stop: while Rodrigues et al. validate via literature database lookups and Sharma et al. via computational rules, we provide a full computational viability assessment comprising predicted synthetic accessibility ($100\%$ of candidates solved by AiZynthFinder), thermodynamic stability (GFN2-xTB) and ADMET profiling, demonstrating the model's utility for prospecting novel, synthetically tractable olfactory candidates.

\section*{Perspective and conclusions}

We have presented an integrated VAE-QSAR framework that addresses the specific data scarcity challenges of olfactory science by combining property prediction with a self-learning generative architecture. By leveraging large-scale chemical data (ChEMBL) to learn molecular syntax and refining the latent space with a specialized odor-prediction objective, the model successfully navigates beyond the boundaries of known odorants. The resulting architecture captures essential structural features while enabling the probabilistic exploration of novel chemical entities that remain physically realizable.

\paragraph{Validation of the Discovery Pipeline:}A critical limitation in generative molecular design is the ``realism gap'' between theoretical validity and experimental feasibility. We addressed this by extending validation beyond the heuristic ``fragrance-likeness'' rules often employed in the field. The model's efficacy is supported by four pillars of evidence:
\begin{enumerate}
    \item \textbf{Synthesizability:} Automated retrosynthetic analysis (AiZynthFinder) identified valid synthesis routes for 100\% of candidates, with an average of 2.89 steps from commercially available precursors, indicating that generated novel scaffolds remain chemically accessible.
    \item \textbf{Stability and Safety:} Quantum mechanical calculations (GFN2-xTB) verified that generated molecules occupy a thermodynamically stable region of chemical space, while ADMET profiling suggests a profile comparable to known commercial odorants on the endpoints assessed, with reactive chemotypes removed by separate structural screening.
    \item \textbf{Generalization:} The model successfully generalized to the external `Unique Good Scents' dataset (unseen during training). This confirms that the decision boundaries learned via SMOTE were biologically relevant rather than artifacts of oversampling, as evidenced by the alignment of the generated distribution with the ground truth manifold (FCD $\approx$ 6.96).
    \item \textbf{Attribution:} An ablation removing the odor prediction head, seeded and sampled identically, yields $4.3$-fold fewer distinct novel compounds per seed and more than doubles outright memorization, while leaving the odor-likeness of the compounds it does produce unchanged. The property loss therefore acts on the \emph{navigability} of the odorous region rather than on the odor-likeness of any individual point within it, and it is this that makes a library of the size analyzed here obtainable at all.
\end{enumerate}

\paragraph{Novelty and predictive reliability:}
A generative framework guided by a
predictive model faces an inherent tension: the predictor is most reliable in the region of
chemical space it was trained on, which is precisely the region a generative model is intended
to leave. Our applicability-domain analysis quantifies this directly, with domain coverage
falling monotonically as structural novelty rises. Two features of the present framework
mitigate it. First, generation is seeded from known odorants and explores outward, so the model
operates by design in the region where the QSAR retains predictive validity; the domain
threshold makes that constraint explicit rather than implicit. Second, the QSAR model acts as a
teacher during training rather than as an oracle certifying individual compounds. The evidence
that the generated library occupies odorant chemical space is independent of the QSAR prediction
for any individual molecule, resting instead on its distributional alignment with verified
odorants, its recovery of odorant frameworks never presented during training, and its
physicochemical and thermodynamic profiles. The compounds reported here are accordingly candidates for experimental
evaluation rather than verified odorants.

\paragraph{Limitations:}
Several constraints determine what the present framework delivers.
The QSAR endpoint is odorousness rather than specific odor qualities: the model predicts whether a compound
is likely to be perceptible as an odorant, not what it will smell like, and its labels do not
distinguish activation of olfactory receptors from trigeminal chemesthesis. Predicted odor
probability is not a measure of prediction reliability, being essentially uncorrelated with
applicability-domain membership, and domain coverage itself declines as structural novelty rises,
so that the compounds which are most novel are those for which the predictor is least
authoritative. Synthetic accessibility is established computationally: retrosynthetic planning
proposes routes from commercially available precursors for all candidates, and their
experimental synthesis is the natural next step, which we intend to pursue.

We note one consequence of the endpoint definition in particular. A prediction of odorousness is
not a prediction of safety. The labels from which the model learns identify compounds as
detectable by smell, which includes reactive and irritant species, and generated compounds
reflect that. We therefore apply a structural filter for reactive chemotypes, removing 360
compounds (1.9\% of the 18,992 candidates at $P(\text{qsar})>0.99$), and report predicted ADMET profiles; neither is a substitute
for the toxicological assessment that would ordinarily precede handling or organoleptic
evaluation of any uncharacterized compound. The compounds reported here are candidates for
experimental evaluation, and their olfactory properties remain predictions.

\paragraph{Future Directions:} While the current implementation predicts only odorousness, and therefore
distinguishes neither odor quality nor olfactory from trigeminal activation, the structured
latent space provides a foundation for fine-grained perceptual design. By conditioning the generative process on multi-label models (predicting specific descriptors like `woody' or `citrus'), researchers could perform targeted inverse design, identifying regions of the latent space associated with complex organoleptic profiles. Furthermore, the confirmation of high synthetic accessibility ($>70\%$ precursor availability) opens the door for accelerated closed-loop discovery cycles. Future work may integrate this generator with automated synthesis and high-throughput screening, using experimental feedback to iteratively refine the latent space mapping. This transition from purely computational generation to a de-risked, experimentally validatable pipeline represents a significant step toward systematically exploring the vast, unmapped regions of olfactory chemical space.

\section*{Acknowledgments}
This research used the ALICE High Performance Computing facility at the University of Leicester.
AI was in receipt of a Wellcome Trust Biomedical Vacation Scholarship.

\section*{Author contribution statement} 
Conceptualization: Tim C. Pearce, Methodology: Ahmed Ibrahim, Tim C. Pearce. Programming: Ahmed Ibrahim, Tim C. Pearce, Formal Analysis: Ahmed Ibrahim, Tim C. Pearce, Investigation: Ahmed Ibrahim, Tim C. Pearce. Writing – Tim C. Pearce.  Visualization: Ahmed Ibrahim, Tim C. Pearce.

{\section*{Code and data availability}
The source code for the Chemical Variational Autoencoder (VAE), trained model weights, and inference pipeline are
publicly available at  \url{https://github.com/tcpearce/odor-molecule-vae/tree/v1.0}. The repository includes: (i) the 1,615 odorant seed molecules used to initialize latent space exploration (\texttt{data/odor\_seeds.csv}), and (ii) the complete set of 37,163 unique generated molecules with their computed physicochemical properties and QSAR-predicted odor probabilities (\texttt{data/generated\_molecules.csv}). Five Supplementary Data files accompany this article: Supplementary Data 1, the full generated library with per-molecule scaffold classification ($n=33{,}507$); Supplementary Data 2, the library at $P(\text{qsar})>0.99$ with per-molecule scaffold classification ($n=18{,}992$; Supplementary Table~S1); Supplementary Data 3 and 4, per-molecule classifications of the ablation arms with and without the QSAR property head (Supplementary Table~S3); and Supplementary Data 5, per-molecule nearest-neighbor Tanimoto similarity to each reference set (Supplementary Table~S4).}

\section*{Funding}
Ahmed Ibrahim was supported by a Wellcome Trust Biomedical Vacation Scholarship. The funders had no role in study design, data
collection and analysis, decision to publish, or preparation of the manuscript


\begin{thebibliography}{99}
\bibitem{1} C.S. Sell, Fundamentals of Fragrance Chemistry, Wiley-VCH, Weinheim, 2019.
\bibitem{2} L. Naldi, Assessment of the risk of fragrance allergy in the general population: challenges and methodological issues, Drug Saf 31 (2008) 440–443. \url{https://doi.org/10.2165/00002018-200831050-00012}.
\bibitem{3} F. Michailidou, The scent of change: sustainable fragrances through industrial biotechnology, ChemBioChem 24 (2023) e202300309. \url{https://doi.org/10.1002/cbic.202300309}.
\bibitem{4} K.A.D. Swift, P. Kraft, eds., Perspectives in flavor and fragrance research, Royal Society of Chemistry (Great Britain), Society of Chemical Industry (Great Britain), Verlag Helvetica Chimica Acta, Zurich, 2005. \url{https://doi.org/10.1002/9783906390475}.
\bibitem{5} K. Touhara, L.B. Vosshall, Sensing odorants and pheromones with chemosensory receptors, Annual Review of Physiology 71 (2009) 307-332. \url{https://doi.org/10.1146/annurev.physiol.010908.163209}.
\bibitem{6} S.M. Kurian, R.G. Naressi, D. Manoel, A.-S. Barwich, B. Malnic, L.R. Saraiva, Odor coding in the mammalian olfactory epithelium, Cell Tissue Res 383 (2021) 445-456. \url{https://doi.org/10.1007/s00441-020-03327-1}.
\bibitem{7} C. Bushdid, M.O. Magnasco, L.B. Vosshall, A. Keller, Humans can discriminate more than 1 trillion olfactory stimuli, Science 343 (2014) 1370-1372. \url{https://doi.org/10.1126/science.1249168}.
\bibitem{8} A. Zhavoronkov, Y.A. Ivanenkov, A. Aliper, M.S. Veselov, V.A. Aladinskiy, A.V. Aladinskaya, V.A. Terentiev, D.A. Polykovskiy, M.D. Kuznetsov, A. Asadulaev, Y. Volkov, A. Zholus, R.R. Shayakhmetov, A. Zhebrak, L.I. Minaeva, B.A. Zagribelnyy, L.H. Lee, R. Soll, D. Madge, L. Xing, T. Guo, A. Aspuru-Guzik, Deep learning enables rapid identification of potent DDR1 kinase inhibitors, Nat Biotechnol 37 (2019) 1038–1040. \url{https://doi.org/10.1038/s41587-019-0224-x}.
\bibitem{9} C.W. Coley, Defining and exploring chemical spaces, Trends in Chemistry 3 (2021) 133–145. \url{https://doi.org/10.1016/j.trechm.2020.11.004}.
\bibitem{10} J. Meyers, B. Fabian, N. Brown, De novo molecular design and generative models, Drug Discovery Today 26 (2021) 2707–2715. \url{https://doi.org/10.1016/j.drudis.2021.05.019}.
\bibitem{11} B. Sanchez-Lengeling, A. Aspuru-Guzik, Inverse molecular design using machine learning: Generative models for matter engineering, Science 361 (2018) 360–365. \url{https://doi.org/10.1126/science.aat2663}.
\bibitem{12} T. Yoshizawa, S. Ishida, T. Sato, M. Ohta, T. Honma, K. Terayama, A data-driven generative strategy to avoid reward hacking in multi-objective molecular design, Nat Commun 16 (2025) 2409. \url{https://doi.org/10.1038/s41467-025-57582-3}.
\bibitem{13} S. Liu, D. Zhang, Z. Tu, H. Dai, P. Liu, Evaluating molecule synthesizability via retrosynthetic planning and reaction prediction, (2025). \url{https://doi.org/10.48550/arXiv.2411.08306}.
\bibitem{14} E.J. Mayhew, C.J. Arayata, R.C. Gerkin, B.K. Lee, J.M. Magill, L.L. Snyder, K.A. Little, C.W. Yu, J.D. Mainland, Transport features predict if a molecule is odorous, Proceedings of the National Academy of Sciences 119 (2022) e2116576119. \url{https://doi.org/10.1073/pnas.2116576119}.
\bibitem{15} R. Gómez-Bombarelli, J.N. Wei, D. Duvenaud, J.M. Hernández-Lobato, B. Sánchez- Lengeling, D. Sheberla, J. Aguilera-Iparraguirre, T.D. Hirzel, R.P. Adams, A. Aspuru-Guzik, Automatic chemical design using a data-driven continuous representation of molecules, ACS Cent. Sci. 4 (2018) 268–276. \url{https://doi.org/10.1021/acscentsci.7b00572}.
\bibitem{16} D.P. Kingma, M. Welling, Auto-encoding variational Bayes, in: 2nd International Conference on Learning Representations (ICLR), Banff, 2014. \url{https://doi.org/10.48550/arXiv.1312.6114}
\bibitem{17} H. Moriwaki, Y.-S. Tian, N. Kawashita, T. Takagi, Mordred: a molecular descriptor calculator, Journal of Cheminformatics 10 (2018) 4. \url{https://doi.org/10.1186/s13321-018-0258-y}.
\bibitem{18} N.V. Chawla, K.W. Bowyer, L.O. Hall, W.P. Kegelmeyer, SMOTE: synthetic minority over-sampling technique, Journal of Artificial Intelligence Research 16 (2002) 321–357. \url{https://doi.org/10.1613/jair.953}.
\bibitem{19} D.W. Hosmer, S. Lemeshow, R.X. Sturdivant, Applied Logistic Regression, 3. Aufl, Wiley, Hoboken, N.J, 2013.
\bibitem{20} Mordred 1.2.1a1 documentation, (n.d.). \url{https://mordred-descriptor.github.io/documentation/master/} (accessed April 22, 2025).
\bibitem{21} D. Mendez, A. Gaulton, A.P. Bento, J. Chambers, M. De Veij, E. Félix, M.P. Magariños, J.F. Mosquera, P. Mutowo, M. Nowotka, M. Gordillo-Marañón, F. Hunter, L. Junco, G. Mugumbate, M. Rodriguez-Lopez, F. Atkinson, N. Bosc, C.J. Radoux, A. Segura-Cabrera, A. Hersey, A.R. Leach, ChEMBL: towards direct deposition of bioassay data, Nucleic Acids Research 47 (2019) D930–D940. \url{https://doi.org/10.1093/nar/gky1075}.
\bibitem{22} RDKit: Open-source cheminformatics, \url{https://www.rdkit.org} (version 2026.03.3). \url{https://doi.org/10.5281/zenodo.591637}
\bibitem{23} K. Preuer, P. Renz, T. Unterthiner, S. Hochreiter, G. Klambauer, Fréchet ChemNet Distance: A Metric for Generative Models for Molecules in Drug Discovery, Journal of Chemical Information and Modeling 58 (2018) 1736–1741 \url{https://doi.org/10.1021/acs.jcim.8b00234}
\bibitem{24} G.M. Maggiora, On outliers and activity cliffs -- why QSAR often disappoints, Journal of Chemical Information and Modeling 46 (2006) 1535. \url{https://doi.org/10.1021/ci060117s}
\bibitem{25} M. Bezaatpour, M. Dal Maso, M. Rissanen, VaPOrS v1.0.1: an automated model for functional group detection and property prediction of organic compounds via SMILES notation, Geosci. Model Dev., 18 (2025) 9189–9217, \url{https://doi.org/10.5194/gmd-18-9189-2025}
\bibitem{26} K. Swanson, et al., ADMET-AI: a machine learning ADMET platform for evaluation of large-scale chemical libraries, Bioinformatics, 40(12) (2024) btae416.
\url{https://doi.org/10.1093/bioinformatics/btae416}
\bibitem{27} P. Ertl, A. Schuffenhauer, Estimation of synthetic accessibility score of drug-like molecules based on molecular complexity and fragment contributions, Journal of Cheminformatics 1 (2009) 8. \url{https://doi.org/10.1186/1758-2946-1-8}
\bibitem{28} S. Genheden, A. Thakkar, V. Chadimová, J-L. Bernstein, O. Engkvist, E. J. Bjerrum, AiZynthFinder: a fast, robust and flexible open-source software for retrosynthetic planning, Journal of Cheminformatics 12(1) (2020) 35.
\url{https://doi.org/10.26434/chemrxiv.12465371.v1}
\bibitem{29} C. Bannwarth, S. Ehlert, S. Grimme, GFN2-xTB—An accurate and broadly parametrized self-consistent tight-binding quantum chemical method with multipole electrostatics and density-dependent dispersion contributions, Journal of Chemical Theory and Computation 15(3) (2019) 1652-1671.
\url{https://doi.org/10.1021/acs.jctc.8b01176}
\bibitem{30} B.C.L. Rodrigues, V.V. Santana, S. Murins, I.B.R. Nogueira, Molecule Generation and Optimization for Efficient Fragrance Creation, Ind. Eng. Chem. Res. 63 (2024) 14667–14681. \url{https://doi.org/10.1021/acs.iecr.4c00650}
\bibitem{31} M. Sharma, S. Balaji, P. Saha, R. Kumar, Navigating the Fragrance Space Using Graph Generative Models and Predicting Odors. Journal of Chemical Information and Modeling 65(10) (2025) 4818-4832.
\url{http://doi.org/10.1021/acs.jcim.5c00209}
\end{thebibliography}
\end{document}